\documentclass[aps,prf,preprint,amsmath,amssymb,groupedaddress]{revtex4-2}
\usepackage{graphicx}% Include figure files
\usepackage{dcolumn}% Align table columns on decimal point
\usepackage{bm}% bold math
\usepackage[colorlinks,citecolor=blue]{hyperref}
\usepackage{mathrsfs}
\usepackage{subfig}
\usepackage{graphicx}
\usepackage{epstopdf}
\usepackage{natbib}
\usepackage{siunitx}
\usepackage{booktabs}
\usepackage{xcolor}
\usepackage{cleveref}
\begin{document}

% Use the \preprint command to place your local institutional report
% number in the upper righthand corner of the title page in preprint mode.
% Multiple \preprint commands are allowed.
% Use the 'preprintnumbers' class option to override journal defaults
% to display numbers if necessary
%\preprint{}

%Title of paper
%\title{Physics-infused Deep Learning sub-grid scale models for large eddy simulations in turbulent fluid flow}
\title{Accurate deep learning sub-grid scale models for large eddy simulations}

% repeat the \author .. \affiliation  etc. as needed
% \email, \thanks, \homepage, \altaffiliation all apply to the current
% author. Explanatory text should go in the []'s, actual e-mail
% address or url should go in the {}'s for \email and \homepage.
% Please use the appropriate macro foreach each type of information

% \affiliation command applies to all authors since the last
% \affiliation command. The \affiliation command should follow the
% other information
% \affiliation can be followed by \email, \homepage, \thanks as well.

\author{Rikhi Bose}
\email{rikhi.bose@gmail.com}
\affiliation{Max Planck Institute for Solar System Research, G\"{o}ttingen, 37077, Germany}
%\affiliation{Mechanical Engineering, Johns Hopkins University, Baltimore, MD 21218, USA.}
\author{Arunabha M. Roy}
%\email{arunabhr.umich@gmail.com}
\affiliation{Aerospace Engineering, University of Michigan, Ann Arbor, MI 48109} %Materials and Structural Systems Division, 
%\affiliation{Aerospace Engineering, Iowa State University, Ames, IA 50011, USA.}

%Collaboration name if desired (requires use of superscriptaddress
%option in \documentclass). \noaffiliation is required (may also be
%used with the \author command).
%\collaboration can be followed by \email, \homepage, \thanks as well.
%\collaboration{}
%\noaffiliation

\date{\today}

\begin{abstract}

We present two families of sub-grid scale (SGS) turbulence models developed for large-eddy simulation (LES) purposes. 
Their development required the formulation of physics-informed robust and efficient Deep Learning (DL) algorithms which, unlike state-of-the-art analytical modeling techniques can produce high-order complex non-linear relations between inputs and outputs. 
Explicit filtering of data from direct simulations of the canonical channel flow at two friction Reynolds numbers $Re_\tau\approx 395$ and 590 provided accurate data for training and testing. 
The two sets of models use different network architectures. 
One of the architectures uses tensor basis neural networks (TBNN) and embeds the simplified analytical model form of the general effective-viscosity hypothesis \citep{lund1993parameterization}, thus incorporating the Galilean, rotational and reflectional invariances. 
The other architecture is that of a relatively simple network, that is able to incorporate the Galilean invariance only. 
However, this simpler architecture has better feature extraction capacity owing to its ability to establish relations between and extract information from cross-components of the integrity basis tensors and the SGS stresses. 
Both sets of models are used to predict the SGS stresses for feature datasets generated with different filter widths, and at different Reynolds numbers. 
It is shown that due to the simpler model's better feature learning capabilities, it outperforms the invariance embedded model in statistical performance metrics. 
In a priori tests, both sets of models provide similar levels of dissipation and backscatter. 
Based on the test results, both sets of models should be usable in a posteriori actual LESs. 
%Relatively improved performance of the invariance embedded models away from the wall indicates that these models are more suitable for modeling isotropic turbulence

\end{abstract}

% insert suggested PACS numbers in braces on next line
% insert suggested keywords - APS authors don't need to do this

%\maketitle must follow title, authors, abstract, \pacs, and \keywords
\maketitle

% \begin{figure}
% \includegraphics{}%
% \caption{\label{}}
% \end{figure}

% Surround figure environment with turnpage environment for landscape
% figure
% \begin{turnpage}
% \begin{figure}
% \includegraphics{}%
% \caption{\label{}}
% \end{figure}
% \end{turnpage}

\section{Introduction}
\label{intro}

Recent rapid developments in the Deep Learning (DL) algorithms \citep{lecun2015deep} and associated softwares \citep{abadi2016tensorflow} have resulted in unprecedented advancements in the use of DL in various scientific disciplines, such as but not restricted to, computer vision \citep{mohr2014computer, roy2022fast}, object detection \citep{menezes2023continual}, image/ signal classification \citep{cao2016extreme}, brain–computer interfaces \citep{li2023novel}, time-series forecasting \citep{bose2022real}, extreme-event prediction \citep{bose2023simulation}. 
The most significant property of DL algorithms is their ability to learn intricate relations between the outputs and the input features. 
This property is enhanced by the use of large volumes of good quality data. 
Therefore, it can be particularly useful in such areas as Computational Fluid Dynamics (CFD), where big volumes of data are available, which enables the excavation of complex high-dimensional relationships. 
Turbulence is the most interesting problem in fluid dynamics as it involves complicated non-linear spatial and temporal evolution of seemingly chaotic flow features, even though its governing partial differential equations are deterministic. 
Additionally, turbulence involves the interaction of a very broad range of spatial and temporal scales, whose adequate resolution is important for correctly capturing the physical phenomenon. 
Direct numerical simulation (DNS) \citep{moin1998direct} or large-eddy simulation (LES) \citep{lesieur1996new} methodologies have been utilized over the past decades to generate huge volumes of data in an attempt to understand this phenomenon. 
%Although much has been learnt about the physical processes involved, 
The available data also provides unprecedented opportunities for modeling turbulence. %exploration of the physical phenomenon, and its modeling. 
THe meteoric rise of the DL algorithms and their easy availability and use have led to widespread efforts to explore their possibilities (see \citet{duraisamy2019turbulence} for a review). 

\subsection{Sub-grid scale (SGS) modeling}

In most geophysical and engineering fluid flow configurations the governing parameter that determines the relative effects of inertia and viscous forces, called the Reynolds number ($Re$), has very large values. 
For example, turbulence in the atmospheric boundary layers corresponds to $Re \sim 10^7$, on an airplane at cruise $Re\sim 10^6 $--$10^7$. 
The computational requirements for resolving all possible length and time scales for any flow is of the order of $\sim Re^{11/4}$ \citep{durbin2011statistical}, and exceed the capabilities of even the most powerful supercomputers. 
DNS resolves all possible spatial and temporal scales and is therefore only viable for relatively low-$Re$ simulations in simple flow configurations. 
In an alternative approach for moderate to high-$Re$ flows, only large-scale motions are resolved, while the effects of small-scale turbulence are modelled \citep{sagaut2005large, meneveau2000scale}. 
In such Large Eddy Simulations (LES), sub-grid scale (SGS) turbulence models represent the physical effects of small-scale turbulence below the grid scale. 
A low-pass spatial filtering decomposes the total flow field into the grid-scale and SGS flows. 
%Because of the decomposition of the total flow field into grid-scale and SGS flow ( associated with the filtering operation at the grid-scale, the total flow is decomposed into the grid-scale and SGS flow components. 
There is a two-way interaction between the resolved flow and its SGS counterpart; in the forward cascading, the grid-scale flow `dissipates' energy to the smaller scale SGS components. 
In the back-scattering mechanism, SGS flow may provide energy to the large-scale flow. 
The objective of this paper is to present the formulation of simple, yet robust, and accurate SGS models for LES of channel flow by leveraging a data-based DL approach within an analytical model-informed framework.

Over the years, the principles of the SGS turbulence modeling have been guided by the observations made on small-scale turbulence physics in the inertial range. 
Although physics-based modeling has been preferred, no universally applicable model has been developed thus far; even the most popular models are inadequate in one way or another. 
Among the existing models, the eddy viscosity models are most widely used. 
In a general SGS eddy viscosity model, the Boussinesq hypothesis is used relating the SGS stress tensor $\tilde{\boldsymbol{\tau}}$ and the rate of strain resolved by the grid, $\tilde{\boldsymbol{S}}=\frac{1}{2}(\nabla\boldsymbol{\tilde{u}} + \nabla\boldsymbol{\tilde{u}}^T)$, linearly, so that, $\tilde{\boldsymbol{\tau}} = \nu_{SGS} \tilde{\boldsymbol{S}}$. 
Here, $\nu_{SGS}$ is the SGS eddy viscosity, generally expressed in the form $\nu_{SGS} \approx C \Delta ^ 2 f(\nabla\tilde{\boldsymbol{u}})$. 
Here, $\Delta$ is a length scale generally taken based on the local grid and $f(\nabla\tilde{\boldsymbol{u}})$ is a functional of $\nabla\tilde{\boldsymbol{u}}$. 
\citet{smagorinsky1963general} first proposed an eddy-viscosity SGS model in which, $f(|\tilde{\boldsymbol{S}}|)$. 
Other eddy-viscosity SGS models formulated based on different functionals $f$ have also been formulated \citep{kim1995new, nicoud1999subgrid, vreman2004eddy, verstappen2011does, trias2015building, silvis2017physical}. 
Although the eddy-viscosity SGS models are able to incorporate the grid-scale to SGS forward energy cascading by dissipating energy from the grid-scale flow, these models are  not capable of providing any SGS to grid-scale energy transfer via back scattering. 
However, the dynamic variants of the eddy viscosity models can locally incorporate back-scattering by predicting negative $\nu_{SGS}$. 
The coefficient $C$ in the expression for $\nu_{SGS}$ above is dynamically updated both spatially and temporally in dynamic models. 
The dynamic models (e.g., \citep{germano1991dynamic, lilly1992proposed, kim1995new, meneveau1996lagrangian}) have been found to provide more accurate predictions, especially when complex flow attributes are involved, such as the influence of pressure gradients, flow separation and reattachment \cite{bose2021simulations, bose2022, bose2023effect}, etc. 
However, the back-scattering is often required to be reduced or entirely eliminated in an LES for the sake of achieving numerical stability. 
%This is achieved by local/ global/ temporal averaging of the computed coefficients. 
The Scale-similarity models (SSM) also demonstrate superior back-scattering properties \citep{bardina1980improved, liu1994properties, domaradzki1997subgrid}. 
\citet{liu1994properties} showed that the correlation coefficients computed between the predictions of the SSM and the true SGS stresses computed from DNS are substantially higher than those used in the eddy-viscosity models. 
In an SSM, the SGS stresses are approximated by the Leonard stresses \citep{leonard1975energy} that are computed by further low-pass filtering the grid-scale flow. 
However, the SSMs are not capable of adding enough dissipation for the grid-scale flow. 
To circumvent this issue, in mixed models \citep{zang1993dynamic, vreman1994formulation, salvetti1995priori, sarghini1999scale}, an SSM is used in conjunction with a purely dissipative constant/ dynamic coefficient eddy-viscosity model. 
%Other types of models include the gradient based models \citep{clark1979evaluation, liu1994properties}. 

\subsection{DL techniques applied to SGS modeling}

NNs have recently been deployed for the purpose of SGS modelling of isotropic turbulence with some success \cite{vollant2017subgrid, wang2018investigations, maulik2019subgrid, zhou2019subgrid, beck2019deep, xie2020modeling, xie2020artificial}. 
Both incompressible and compressible cases have been explored. 
In general, the NNs have been shown to performance as well if not better than the analytical models. 
The fully connected neural networks (FCNN) have been preferred, although \citet{beck2019deep} showed that for the case of 3-D decaying homogeneous isotropic turbulence, a convolutional neural network (CNN) \citep{lecun1995convolutional} predicted the SGS forces better than the FCNNs. 

In comparison, the use of the DL techniques in inhomogeneous wall-bounded turbulence has been less explored. 
\citet{sarghini2003neural} trained FCNNs with input features generated from the LES of a low-$Re$ channel flow simulation to predict the model coefficient in an eddy-viscosity turbulence model. 
Later, \citet{wollblad2008pod} applied Proper Orthogonal Decomposition (POD) to decompose the SGS stresses obtained from filtered DNS data, and FCNNs were trained on input features generated from filtered flow fields to predict the coefficients of the truncated series. 
In a priori tests, the model predicted the SGS stresses reasonably well. 
However, in a posteriori tests, they had to use a linear combination of the SGS stress predictions from an FCNN model and the Smagorinsky eddy-viscosity SGS model for the sake of stabilizing their flow solver. 
\citet{gamahara2017searching} applied six different FCNNs to predict the six independent components of the symmetric SGS stress tensor, $\tilde{\boldsymbol{\tau}}$. 
The FCNNs were trained on four sets of input features generated by filtering DNS flow fields, such as, the components of the filtered velocity gradient tensor ($\nabla \tilde{\boldsymbol{u}}$), $\tilde{\boldsymbol{S}}$, and of the rotation rate tensor, $\tilde{\boldsymbol{R}}= \frac{1}{2}(\nabla\boldsymbol{\tilde{u}} - \nabla\boldsymbol{\tilde{u}}^T)$. 
The models performed reasonably well in both a priori and a posteriori tests. 
However, because of the use of the six independent NNs, the correlations between the six independent components were not maintained. 
\citet{pal2020deep} also used the input features from LES to train FCNNs for the purpose of predicting the eddy-viscosity coefficients generated by the dynamic Smagorinsky model. 
Predictions by the FCNNs on coarser grids were shown to be better than those based on the dynamic Smagorinsky model. 
The approaches by \citet{sarghini2003neural} and \citet{pal2020deep} yielded eddy-viscosity type DL-SGS models, and the input features were not Galilean invariant. 
\citet{park2021toward} used FCNNs to predict the SGS stresses in both a priori and a posteriori tests. 
Unlike \citet{gamahara2017searching}, they used one network to predict the components of $\tilde{\boldsymbol{\tau}}$ in turbulent channel flow simulations. 
However, their input features were similar to those used by \citet{gamahara2017searching}, i.e., components of the $\nabla\tilde{\boldsymbol{u}}$, $\tilde{\boldsymbol{S}}$, $\boldsymbol{u}$, and $\frac{\partial \tilde{\boldsymbol{u}}}{\partial  y}$ (wall-normal gradient of velocity). 
Although the trained models performed reasonably well, in the conclusions they commented, `$\ldots$ one may also consider other combinations of $\tilde{S}_{ij}$ and $\tilde{R}_{ij}$ as input variables, $\ldots$ Thus, a further study in this direction is needed.'  
The present work explores this direction within an analytical modeling framework. 
%within the framework of the general effective viscosity hypothesis first proposed by \citet{pope1975more}. 

\subsection{Motivation}
The modeling of turbulence requires consideration of the physical information provided by the governing equations. 
Consequently, the use of the DL algorithms for turbulence modeling is significantly different from their use in other fields. 
For example, in incompressible fluid turbulence, the instantaneous velocity field must be divergence free, and this constraint must be satisfied at each time instant. 
Experimentalists often reconstruct the full flow fields from sparse observations by making the observations explicitly adhere to the solenoidality constraint imposed on the velocity \citep{gesemann2016noisy}. 
In conventional CFD solvers based on numerical discretization of the governing equations, this condition is generally satisfied by solving the Poisson equation for pressure in addition to the momentum equations for velocity. 
In a data-driven approach, the momentum solvers must be developed so that they satisfy the solenoidality condition for the velocity field, therefore, restricting the solution space (see e.g., \citet{kochkov2021machine, list2022learned}). 
The other constraint that is specific to turbulence modeling is that, a formulated model should be independent of the coordinate system, i.e., invariant to the translation (Galilean invariance), and rotation or reflection about an axis \citep{durbin2011statistical}. 
In modeling terms utilizing DL algorithms, the model must be consistent and not change if the models are tested on coordinate systems that may be translated, rotated or reflected about an arbitrary axis different from the coordinate system on which the input and output features are generated for the purpose of training \citep{wang2017physics, wu2018physics}.

The SGS stresses are intrinsically invariant under Galilean transformation \citep{hartel1997galilean}. 
\citet{speziale1985galilean} showed that some of the filtering techniques other tan the sharp spectral cutoff filter do not satisfy the Galilean invariance property for different components of the SGS stresses. 
Additionally, the SGS stresses also satisfy the rotational and reflectional invariance in case the filter function is isotropic in nature. 
In wall-bounded turbulence, because of the anisotropy induced by the presence of the wall, the applied filter function cannot be isotropic, and therefore, the SGS stresses also do not satisfy the rotational and reflectional invariance properties in the vicinity of the wall. 
Therefore, the DL-SGS models need not have these properties locally. 
\citet{ling2016reynolds} showed that the embedding of the Galilean invariance can significantly improve the DL model performance in predicting Reynolds stresses in Reynolds-averaged Navier Stokes (RANS) simulations. 
On the other hand, the literature on DL-SGS models for wall bounded turbulence \citep{gamahara2017searching, pal2020deep, park2021toward}, the Galilean invariance property of the SGS stresses were not taken into account. 
Specifically, the input features did not embed either of the necessary invariance properties. 
In the present paper, we attempt to alleviate these limitations of the state-of-the-art approaches. 
We attempt to formulate a NN architecture for wall-bounded SGS modeling that embeds all necessary invariance properties everywhere in the domain by emulating an analytical model form in a DL framework. 
Furthermore, careful choice of filter function ensures automatic satisfaction of the Galilean invariance property of the predicted SGS stresses. 
Additionally, we introduce a simpler yet effective family of NN models specifically designed for wall-bounded SGS turbulence modeling. %that can substantially outperform the invariance-embedded DL-SGS models. 

\subsection{Contribution}
Utilizing the powerful feature extraction capabilities of DL algorithms, we present two families of robust and accurate SGS turbulence models. 
One of the modeling approaches takes into consideration an analytical expansion of the SGS turbulence stresses into integrity basis tensors composed of the symmetric and anti-symmetric parts of the resolved velocity gradient tensor ($\tilde{\boldsymbol{S}}$ and $\tilde{\boldsymbol{R}}$, respectively) \citep{pope1975more, lund1993parameterization}. 
To this end, we leverage the tensor-basis neural network (TBNN) \citep{ling2016reynolds} for the first time in the context of SGS turbulence modeling in order to in a DL model architecture the form of the analytical expansion proposed by \citet{lund1993parameterization}. 
This family of models has two input layers, an invariant input layer, and a tensor input layer. 
DL representation of the analytical model form helps us achieve invariance under Galilean, rotational and reflectional transformations for this family of NN models. 
From here on, we call this approach $A1$. 

In the second, more data-driven approach, called $A2$, a structurally simpler NN also uses the independent components of all the integrity basis tensors in addition to the invariant inputs in a single input layer unlike the TBNN. 
%used in the invariant input layer of the TBNN architecture 
This second approach is motivated by the fact, that in wall-bounded turbulence, the wall provides an automatic choice of the reference frame, and therefore, embedding of all the invariance properties may not be necessary for this case-specific situation. 
Hence, we relax the invariance requirements, and find that, NN models developed embedding only the Galilean invariance property, outperform the models with all the invariance properties embedded in them. 
%This approach is named $A2$. 
We demonstrate that $A2$ (the more data-driven approach) is more efficient at predicting the SGS stresses despite using a simpler NN architecture. 
It is noted that even the more data-driven approach, $A2$ incorporates physical information to some extent as the input features are also inspired by the same analytical model of turbulence as used in the approach $A1$.

\section{Data generation}
\label{sec2}

\subsection{Direct numerical simulations (DNS)}

\begin{figure}[!ht]
    \begin{center}
        \includegraphics[trim=0 0 0 0,clip,width=0.85\textwidth]{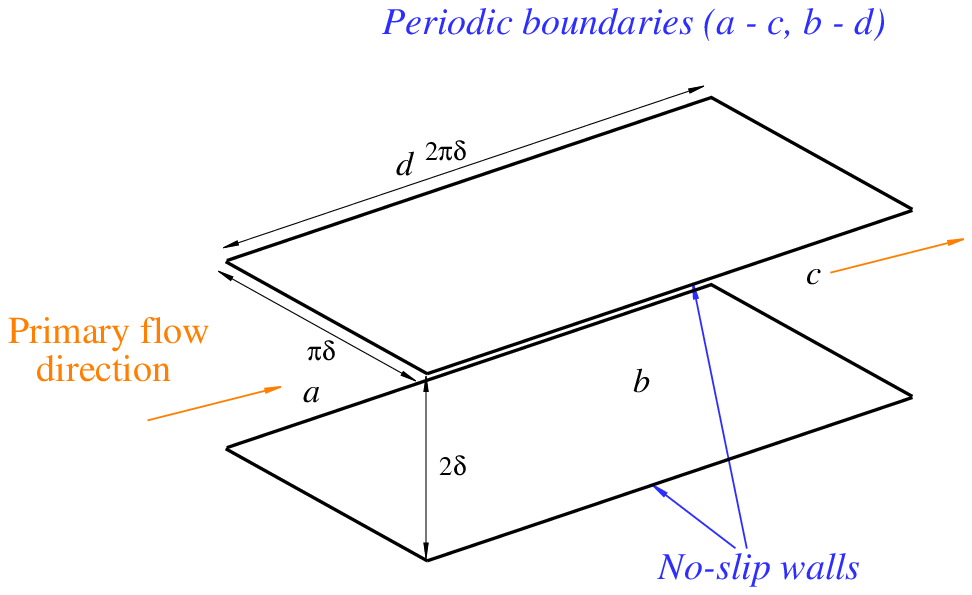}
        \caption{Schematic of the flow configuration.}
        \label{f1}
    \end{center}
\end{figure}

In the first stage of the work, direct simulations are performed for the canonical channel flow problem. 
In the problem being considered, the primary flow is bounded by two plates, one at the top, and the other at the bottom (see Fig. \ref{f1}). 
Theoretically, the plates are infinite horizontally. 
In our simulations, the dimension of the plates are $2\pi \delta \times \pi \delta$ in the along-stream and across-stream (span-wise) directions; the two plates are $2\delta$ apart. 
No-slip boundary conditions are applied at the plate walls. 
The planes normal to the plate boundaries are named $a$--$d$ in Fig. \ref{f1}. 
To accommodate three-dimensionality, periodic boundary conditions are applied at these boundaries: planes ($a$, $c$), and ($b$, $d$).

In the direct numerical simulations (DNS), the incompressible Navier-Stokes equations are solved. 
The solenoidality condition applied to the velocity field ($u_i$) as in Eq. \ref{dns1} ensures incompressibility of the flow field. 

\begin{eqnarray}
  \label{dns1}
  & \partial_i u_i = 0 \\
  & \partial_t u_i + \partial_j (u_iu_j) = -\frac{1}{\rho}\partial_i p + \nu\partial_j^2 u_i
  \label{dns2}
\end{eqnarray} 
Here, directions 1, 2, and 3 represent the stream-wise ($x$), wall-normal ($y$) and span-wise ($z$) directions, respectively. 
The discretized equations are solved on a staggered grid with pressure at the cell center. 
The flow solver is based on a second-order accurate finite difference discretization. 
A semi-implicit time-marching method advances the velocity field in time. 
The equation derived for pressure correction is Fourier transformed in the horizontal directions; consequently a system of linear equations are solved corresponding to each wavenumber in the wall-normal direction to obtain the pressure correction term. 
A divergence-free velocity field is obtained at the new time step following application of the pressure correction. 
A constant body force was applied at each computational cell to obtain a flow at the desired Reynolds number based on friction.

\begin{table}
	\centering
	\begin{tabular*}{\linewidth}{@{\extracolsep{\fill}}c c c c c }
		%\hline
		%\\[-0.5em]
		\hline
		%\\[-0.8em] % Adds extra space after hline
		$Re_\tau$			& Domain size           & Grid                        & Horizontal grid 		&  Vertical grid
		\\
		  ($\frac{u^* \delta}{\nu}$)	& ($x\times y\times z$) & ($N_x\times N_y\times N_z$) &   ($\Delta x^+\times \Delta z^+$) &  ($\Delta y^+_{max}$)
		\\[-0.0em]
		\hline
		% 		\\[-0.1em]
		% 		YOLOv2 & 0.831 & 0.827 & ? & 0.783
		% 		\\
		%\\[-0.5em]
		395 & $2\pi\delta \times 2\delta \times \pi\delta$ & $256\times 193 \times 192$ & $9.7 \times 6.4$ & 6.4 
		\\
		%\\[-0.5em]
		590 & $2\pi\delta \times 2\delta \times \pi\delta$ & $384\times 257 \times 384$ &  $9.6 \times 4.8$ & 7.2 
		\\
		%\\[-0.5em]
		\hline
	\end{tabular*}
	\caption{Description of the DNSs performed to generate data}
	\label{t1}
\end{table}

DNSs were performed at two Reynolds numbers. 
The details of the grids for each of these DNSs in viscous wall unit (the velocity scale is the friction velocity $u^*$, the length scale is given by $\nu/u^*$, where $\nu$ is the molecular viscosity) are tabulated in Tab. \ref{t1}. 
The grid spacings are appropriate for capturing the whole turbulence spectrum up to the dissipative Kolmogorov scales. 

\begin{figure}[!ht]
    \begin{center}
        \includegraphics[trim=0 0 0 0,clip,width=0.85\textwidth]{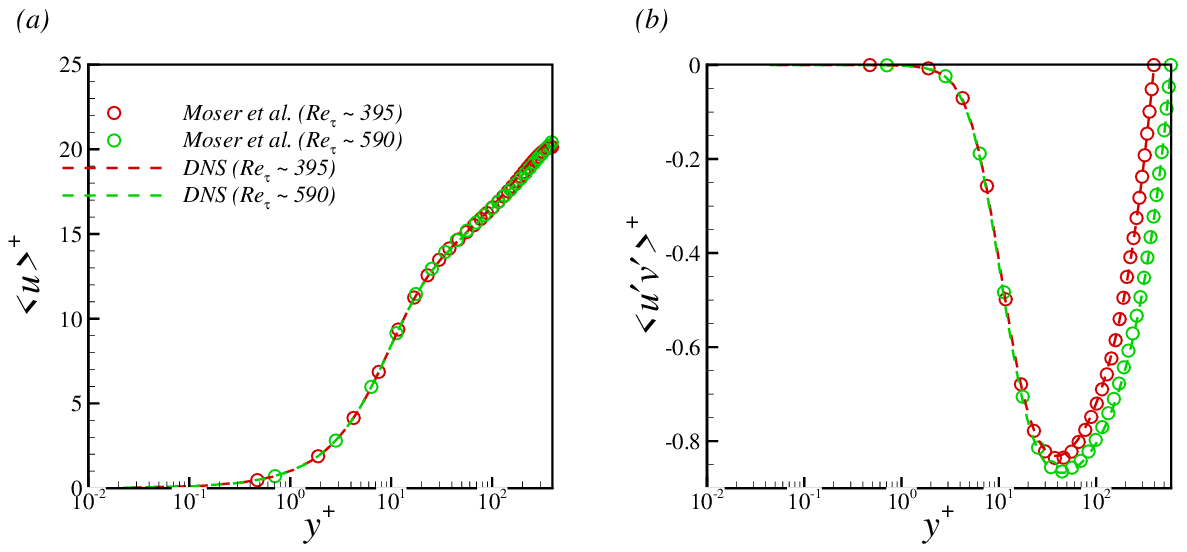}
        \caption{Comparisons of the mean profiles from DNSs listed in Tab. \ref{t1} (dashed lines) with those from \citet{moser1999direct} (symbols): $(a)$ stream-wise velocity $\langle u\rangle^+$ $(b)$ cross-stream ($\langle u'v' \rangle^+$) component of the Reynolds stresses. 
        All variables are shown in the viscous wall unit. 
        }
        \label{fdns1}
    \end{center}
\end{figure} 

\begin{figure}[!ht]
    \begin{center}
        \includegraphics[trim=0 0.1cm 0.1cm 0,clip,width=0.85\textwidth]{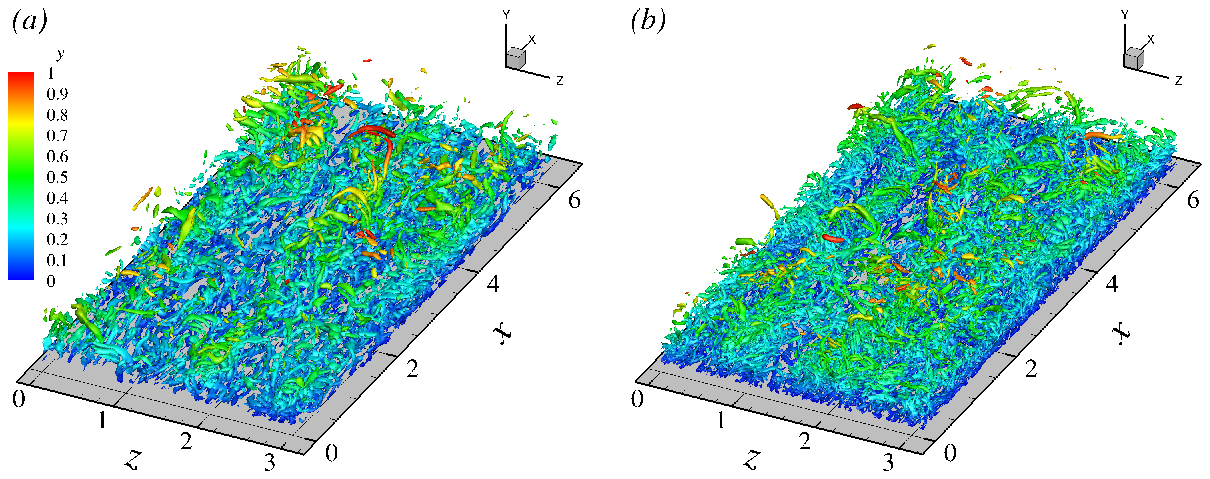}
        \caption{Isosurfaces of the $Q$--criterion depicting the vortical structures: ($a$) at $Re_\tau \sim 395$, and ($b$) $Re_\tau \sim 590$. 
        }
        \label{fdns2}
    \end{center}
\end{figure}

The accuracy of the DNSs are shown by plotting the profile of the mean stream-wise velocity ($\langle u\rangle^+=\langle u\rangle/u^*$) in the wall-normal direction scaled by the viscous wall unit $y+=\frac{u^* y}{\nu}$ in Fig. \ref{fdns1}($a$). 
Henceforth, $\langle \cdot \rangle$ indicates averaging in the horizontal directions as well as in time. 
The non-dimensional cross-stream Reynolds stress ($\langle u'v'\rangle^+$) is also shown in Fig. \ref{fdns1}($b$). 
At both Reynolds numbers, both profiles are in very good agreement with the literature \cite{moser1999direct}. 
To show a snapshot of the instantaneous flow, isosurfaces of the $Q$-criterion are shown for the $Re_\tau \sim 395$ case in Fig. \ref{fdns2}($a$), and for the $Re_\tau \sim 590$ case in Fig. \ref{fdns2}($b$). 
Here, $Q=\frac{1}{2}(|\boldsymbol{R}|^2 - |\boldsymbol{S}|^2)$, where, $\boldsymbol{R}$ and $\boldsymbol{S}$ are the rotation rate and strain rate tensors, respectively, depicts the vortical structures in a flow. 
The near-wall flow is populated by vortices of different scales. 
More vortices in the near-wall flow are evident for the higher Reynolds number case shown in Fig. \ref{fdns2}($b$). 

Each of these simulations was run for a sufficient time to pass the transients, and for the flow to come to a state of statistical stationarity. 
The flow fields were then recorded every $\sim 3.5$ flow-through time units ($\frac{2\pi\delta}{\max(\langle u\rangle)}$). 
The long time lag between saved time snapshots ensured that the data generated for the training and testing purposes are uncorrelated in time. 
Thus, 120 instantaneous snapshots of the statistically stationary flow field were stored for the purpose of filtering, and for preparing input and output features to be used for model training.

\subsection{Large-eddy simulations (LES): Sub-grid scale modeling}

\iffalse
The small-scale fluctuations from a flow field obtained from DNS/ experiments ($f$) may be filtered out explicitly by the application of a filter function,
$$\tilde{f} = \int f(\boldsymbol{x'})G(\boldsymbol{x}-\boldsymbol{x'})d\boldsymbol{x'}$$
Alternatively, the filtered DNS equations may be solved by modeling the small-scale sub-grid scale flow. 
Such an approach is called the large-eddy simulation (LES). 
In an LES, the small-scale fluctuations are modelled, and the grid-scale flow only consists of the large-scale motions. 
The governing equations for LES may be obtained by applying the same filtering procedure on Eqs. \ref{dns1}--\ref{dns2}. 
\fi

The small-scale fluctuations from a flow field obtained from DNS/ experiments ($f$) may be filtered out explicitly by the application of a filter function $G$. 
In physical space, this operation leads to the filtered variable, 

$$\tilde{f} = \int f(\boldsymbol{x'})G(\boldsymbol{x}-\boldsymbol{x'})d\boldsymbol{x'}$$

The filtering operation can be explicit as above, or, alternatively, in an implicit filtering approach, the filtered DNS equations may be solved in a large-eddy simulation (LES). 
In an LES, the small-scale fluctuations are modelled, and the grid only resolves the large-scale motions. 
%So, the grid-scale flow field in an LES ($\tilde{f}$) may be obtained from the DNS flow field ($f$) as follows, $\tilde{f} = \int f(\boldsymbol{x'})G(\boldsymbol{x}-\boldsymbol{x'})d\boldsymbol{x'}$. 
The governing equations for LES are obtained by applying the filtering procedure mentioned above (Eqs. \ref{dns1}--\ref{dns2}). 

\begin{eqnarray}
  \label{les1}
  & \partial_i \tilde{u}_i = 0 \\
	& \partial_t \tilde{u}_i + \partial_j (\tilde{u}_i\tilde{u}_j) = -\frac{1}{\rho}\partial_i \tilde{p} + \nu\partial_j^2 \tilde{u}_i + \partial_j \tilde{\tau}_{ij}
  \label{les2}
\end{eqnarray}

The boundary conditions are the same as for DNS; the only difference is that the conditions are applied for the filtered variables. 
The above equations must be closed by modeling the sub-grid scale (SGS) stresses, $\tilde{\tau}_{ij} = \tilde{u}_i\tilde{u}_j-\widetilde{u_i u_j}$, which quantify the contributions of the small-scale flow. 
For an incompressible flow, the isotropic component of the tensor $\tilde{\tau}_{ij}$ is absorbed in the pressure gradient term on the RHS of Eq. \ref{les2}. 
Because of the symmetry of $\tilde{\tau}_{ij}$, only six of its components are independent and need to be modelled. 
As we only consider turbulence in the incompressible regime, in the rest of the paper, $\tilde{\tau}_{ij}=\tilde{\tau}_{ij}-\frac{1}{3}\tilde{\tau}_{kk}\delta_{ij}$, refer to the deviatoric component  of the SGS stresses. 
%The objective of the present work is to model this term leveraging a data-based Deep Learning (DL) approach. 
Evidently, the model must be based on the flow fields being solved for (the grid-scale flow field) in an LES, or on variables that are derived from those. 

\iffalse
Despite physics based SGS modeling has been preferred over the years, even the most popular models are inadequate in one way or another. 
Among the existing models, the eddy viscosity models are widely used. 
In a general eddy viscosity model, $\tilde{\tau}_{ij} = \nu_{SGS} \tilde{S}_{ij}$, where, $\tilde{S}_{ij}=\frac{1}{2}(\partial_i \tilde{u}_j + \partial_j \tilde{u}_i)$ is the resolved strain rate tensor. 
In such models, often $\nu_{SGS}$ is of the form $\nu_{SGS} \approx C \Delta ^ 2 f(|\tilde{\boldsymbol{S}}|)$. 
Here, $\Delta$ is a grid length scale and $f(|\tilde{\boldsymbol{S}}|)$ is a functional of $\tilde{S}_{ij}$ as in the Smagorinsky model \cite{smagorinsky1963general} for example. 
The functional has been considered based on other flow-variables such as the sub-grid scale kinetic energy, $k$ for the k-equation model \cite{kim1995new} or the second invariant of the the square of the velocity gradient tensor in the WALE model \cite{nicoud1999subgrid} and so on. 
However, no universally applicable model is known thus far. 
The objective of the paper is to present the formulations of simple, yet robust, and accurate SGS models leveraging a data-based Deep Learning approach. 
\fi

Using Machine/ Deep Learning (ML/ DL) techniques, it could be possible to quantitatively ascertain most important flow attributes from the resolved field (quantities with the $\tilde{}$ symbol over them). 
In the present work, we explore a couple of strategies based on DL for the purpose of modeling $\tilde{\tau}_{ij}$. 
The DL method is capable of establishing complex high-dimensional non-linear functional relations between the output, $\tilde{\tau}_{ij}$, and input flow features, which may not be possible in a conventional framework. 
%including several instantaneous flow variables based on the strain-rate and rotation-rate tensors
In the approaches presented herein, the training fields were generated from the DNS flow fields of the canonical channel flow by explicitly applying the aforementioned filtering procedure. 
%A box filter was applied in all three spatial directions. 
Both input features and output turbulent stresses were generated in this way. 
Next, we trained several DL models with suitable architectures depending on the input features and came up with the best performing models based on their error in validation data. 
Finally, the models that provided the least error in the validation data were used to predict the testing datasets on different grids. 
In this phase, several performance metrics used for measuring the performance of the SGS models are presented herein to gauge the efficacy of these models.

\subsection{Filtered fields: Explicit filtering}

\begin{table}
	\centering
	\begin{tabular*}{\linewidth}{@{\extracolsep{\fill}}c c c c c c c}
		\hline
		$Re_\tau$			& Domain size           &            Grid           & Horizontal grid 		&  Vertical grid & Purpose		&   Nomenclature \\
		  ($\frac{u^* \delta}{\nu}$)	& ($x\times y\times z$) & ($N_x\times N_y\times N_z$) &   ($\Delta x^+\times \Delta z^+$) &  ($\Delta y^+_{max}$) &                 &       
		\\[-0.0em]
		\hline
		395 & $2\pi\delta \times 2\delta \times \pi\delta$ & $36\times 48 \times 36$ & $68.9 \times 34.47$ & 26.4 & Train/ Test	& $D1$\\
        395 & $2\pi\delta \times 2\delta \times \pi\delta$ & $48\times 48 \times 48$ & $51.7 \times 25.85$ & 26.4 & Test	&  $D2$	\\  
		590 & $2\pi\delta \times 2\delta \times \pi\delta$ & $54\times 72 \times 54$ & $68.65 \times 34.32$ & 26.1 & Test	& $D3$	\\
		\hline
	\end{tabular*}
	\caption{LES grid parameters corresponding to the filtered fields (FFs) obtained by filtering the DNS data described in Tab. \ref{t1}. 
    The FFs are used to generate input and output features for training and testing the DL-SGS models.}
	\label{t2}
\end{table}

In the present study, the features used as inputs and the target SGS stresses are all obtained from the three components of the velocity field $\boldsymbol{u}$. 
Therefore, in the first stage of preparing the features, components of the instantaneous velocity fields obtained from DNSs listed in Tab. \ref{t1} were filtered in the three spatial directions at each time instant. 

In the horizontal directions, a Fourier transformation was followed by multiplication of the obtained Fourier coefficients by the spectrally sharp cutoff filter function. 
The choice of the filter function is motivated by the finding of \citet{speziale1985galilean}, who showed that the spectrally sharp cutoff filter produces SGS stresses with its various decomposed parts satisfying the Galilean invariance individually. 
In the wavenumber space, the filter function is given by, 

\begin{equation}
    \hat{G}(\boldsymbol{k}) =
    \begin{cases}
      1 & \text{if $|\boldsymbol{k}| \le \boldsymbol{k_c}$}\\
      0 & \text{ otherwise}
    \end{cases}
    \label{filt}
\end{equation}

In this expression, $\boldsymbol{k}_c$ is the cutoff wavenumber. 
Next, the Fourier coefficients are inverse transformed to obtain the filtered variable in the physical coordinates. 
To facilitate the filtering operation in the wall-normal direction, the wall-normal grid locations are chosen at the Chebyshev nodes. 
The flow field is then transformed into the coefficients of the Chebyshev polynomials in the wall-normal direction. 
The filter function is explicitly applied to the coefficients corresponding to each Chebyshev polynomial. 
The LES grid parameters associated with the explicit filtering operation are tabulated in Tab. \ref{t2}. 
The parameters associated with the explicit filtering operation in each direction may be obtained from this information. 
We use the filtered data to generate input features for training, validating, and testing models. 
Testing data generated from the three tabulated LES grids are named in the last column of Tab. \ref{t2} for easy reference later in the paper.

\begin{figure}[!ht]
    \begin{center}
        \includegraphics[trim=0 0.1cm 0.1cm 0,clip,width=0.95\textwidth]{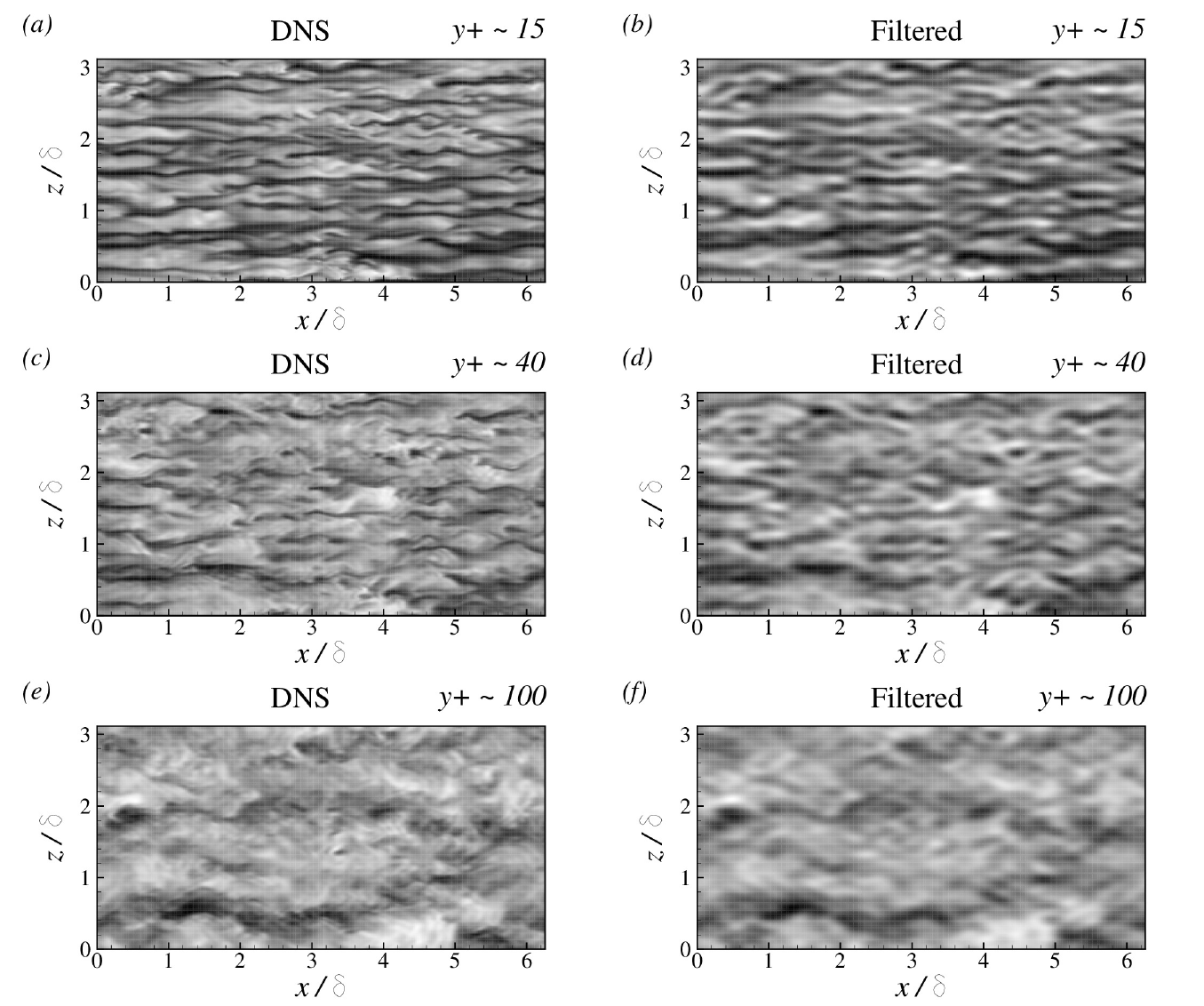}
        \caption{Contours of $|\boldsymbol{u}|/u*$ from DNS (left column) and corresponding filtered field $|\boldsymbol{\tilde{u}}|/u*$ (right column) at different wall-parallel planes for $Re_\tau \sim 395$: $(a)$--$(b)$: $y+ \sim 15$; $(c)$--$(d)$: $y+ \sim 40$, and $(e)$--$(f)$: $y+ \sim 100$ at a chosen time instant. 
        The DNS field is from the simulation listed in Tab. \ref{t1} and the filtered field corresponds to an LES grid: $36\times48\times36$ (see Tab. \ref{t2}). 
        }
        \label{fFilt1}
    \end{center}
\end{figure} 

Instantaneous contours of the scaled magnitude of the velocity ($|\boldsymbol{u}|/u*$) are plotted at various wall-parallel planes from the DNS in the left column of Fig. \ref{fFilt1} and compared with their filtered counterpart ($|\tilde{\boldsymbol{u}}|/u*$) in the right column. 
The three chosen planes are representative of the flow at relevant layers in the wall-normal structure of a wall-bounded turbulent flow field. 
With increasing distance from the wall, the three heights correspond to the buffer layer, the log layer and the outer layer, respectively. 
Close to the wall, the flow is streaky in the stream-wise direction. 
Clearly, the small-scale sharp features are absent in the filtered field (abbreviated as FF from here onwards). 
The flow structures are also streaky at $y+\sim40$, however, small-scale fluctuations are also clearly visible in the DNS field on the left. 
In the outer flow, the flow structures are significantly larger than closer to the wall. 
Evidently, the FFs only retain the large-scale features from the DNS. 
As listed in Tab. \ref{t2}, all models were trained on the features generated from the filtered velocity fields obtained on this grid.

\begin{figure}[!ht]
    \begin{center}
        \includegraphics[trim=0 0.1cm 0.1cm 0,clip,width=0.95\textwidth]{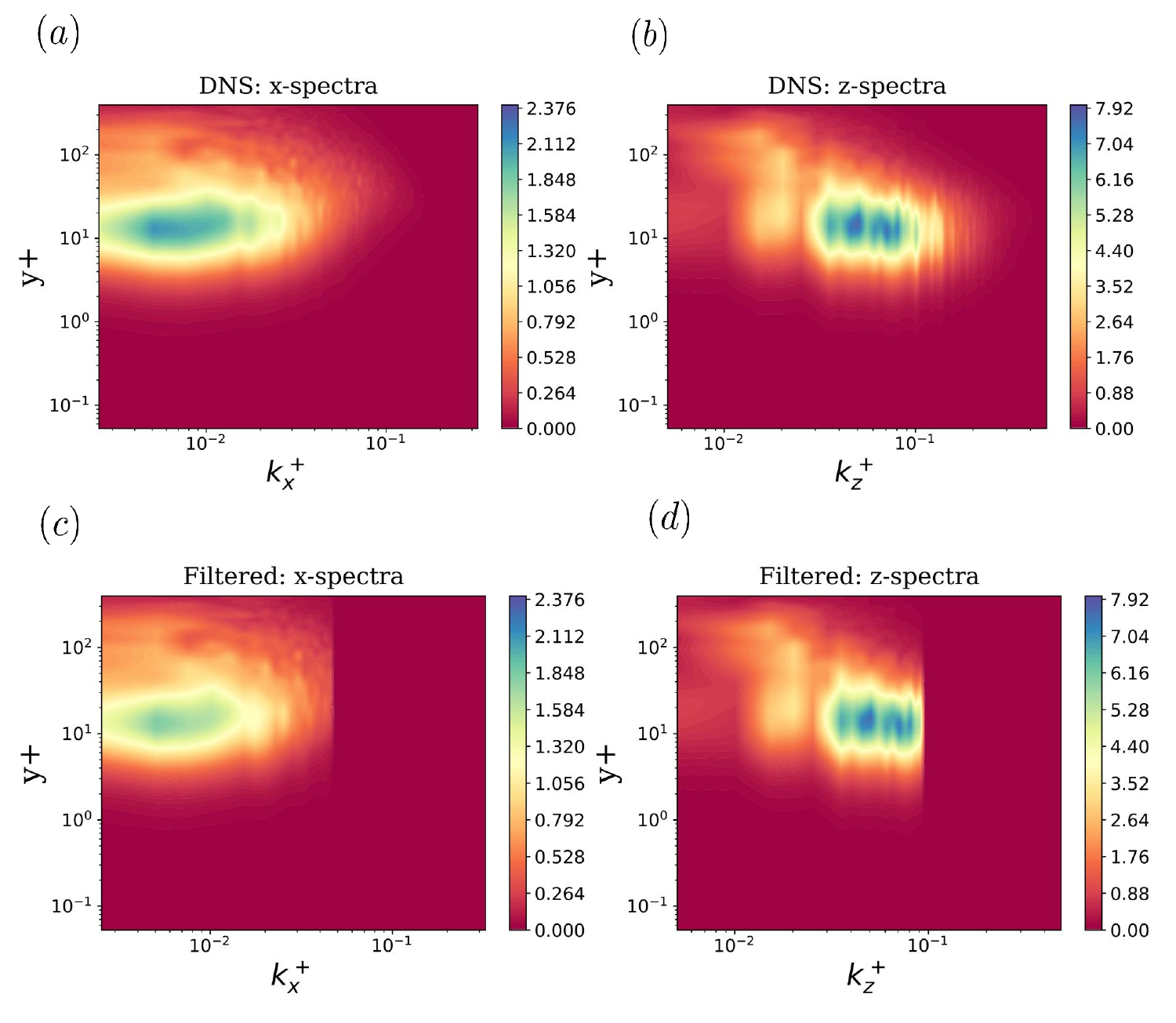}
        \caption{Contours of the scaled stream-wise fluctuation velocity ($u'$) spectra $k_x^+\phi^+_{u'u'}(k_x^+, y^+)$ (left column) and $k_z^+\phi^+_{u'u'}(k_z^+, y^+)$ (right column): $(a)$--$(b)$ from DNS field $u$ at $Re_\tau \sim 395$, and $(c)$--$(d)$ corresponding filtered field $\tilde{u}$. 
        The DNS field is from the simulation listed in Tab. \ref{t1} and the filtered field corresponds to an LES grid: $36\times48\times36$ (see Tab. \ref{t2}). 
        The time instant is the same as that for which flow structures are shown in Fig. \ref{fFilt1}. 
        }
        \label{fFilt2}
    \end{center}
\end{figure}

Scaled pre-multiplied stream-wise wavenumber spectra ($k_x^+\phi^+_{u'u'}(k_x^+, y^+)$) for the stream-wise fluctuation velocity component ($u'$) has been shown from DNS and compared with the FF in the left column of Fig. \ref{fFilt2}. 
The same comparison for the pre-multiplied span-wise wavenumber spectra ($k_z^+\phi^+_{u'u'}(k_z^+, y^+)$) is shown in the right column of Fig. \ref{fFilt2}. 
The spectra are shown at all wall-normal height, and also depict the effect of wall-normal filtering. 
They are shown at the same time instant at which the flow structures are shown in Fig. \ref{fFilt1}. 
The spectra from the DNS are very similar to those reported in the literature, again providing validation for the accuracy of the DNS. 
They clearly show that beyond the cutoff wavenumbers in each of the horizontal directions, the Fourier amplitudes are indeed zero for the FFs due to the nature of the chosen filter function (see Eq. \ref{filt}). 
The peak amplitudes in both the stream-wise and span-wise spectra for the FF are somewhat smaller than the DNS which may be attributed to the effect of wall-normal filtering.

\section{Invariance embedded Neural Networks}
\label{sec3}

Training an ANN generally requires large volumes of data. 
The ANNs are capable of efficiently extracting complex non-linear high-dimensional relations from the data. 
%In general, larger volumes of data contain more useful information repetitively. 
Researchers have often used any possible available information to augment the prediction performance and applicability of the ANN models including large volumes of data. 
The equations governing a system, initial/ boundary conditions have been exploited to constrain the solution space to be learned by the NNs \citep{raissi2019physics, roy2023deep, roy2023data}. 
\citet{ling2016reynolds} utilized the predictive capability of ANNs in conjunction with a generic model for turbulence \citep{pope1975more} to gain predictive capability as well as robustness in model applicability. 
It is therefore essential to carefully design the architecture of the ANNs, the associated training procedure, including the loss functions, input and output features, for the best possible representation of the governing physics, which in the present context is the modeling of the SGS turbulence characteristics. 
In the present work, special care was needed both in the choice of appropriate input features, and in the design of the NN architectures for embedding the necessary invariance properties into the DL-SGS models, so that the developed models are robust and usable in various prediction conditions. 

\subsection{Base Network: FCNN layers}
The fundamental architecture of the trained models is the simple feed-forward fully connected NN. 
In a feed-forward fully connected NN, each neuron in a layer is connected to all the neurons in the immediate neighbour layers; outputs of a neuron in each layer (say layer $n$) is only fed to the neurons in the next layer (i.e., layer $n$ or forward in a directional sense). 
Consequently, for the neurons in the $(n)^{th}$ layer, these are received as the inputs. 
A general expression for such a network is as follows: 

\begin{equation}
     \mathscr{N}^{n}(\mathscr{N}^{n-1}) = \mathscr{\phi}^{n}(\boldsymbol{W}^{n-1}\cdot\mathscr{N}^{n-1} + \boldsymbol{b}^{n-1})
     \label{nn}
\end{equation}

\noindent where, $n \in {1, ..., n_L}$ is the layer number of the ANN. 
The above expression represents a mapping from the input layer (IL) $\mathscr{N}^{0} \in \mathbb{R}^{n=0}$ to the output layer (OL) $\mathscr{N}^{nL} \in \mathbb{R}^{n=nL}$. 
On the right hand side, $\mathscr{\phi}^{n}(\cdot)$ is a non-linear transformer or an activation function that is applied on the vector $(\boldsymbol{W}^{n-1}\cdot\mathscr{N}^{n-1} + \boldsymbol{b}^{n-1})$ element-wise. 
Also, $\boldsymbol{W}^{n} \in \mathbb{R}^{l_n\times l_{n+1}}$ and $\boldsymbol{b}^{n} \in \mathbb{R}^{l_{n}}$ are the weights and biases at the layer $n$. 
On both sides, $\mathscr{N}$ represents the non-linear mapping of the vector $\boldsymbol{y}$ defined as, $$\mathscr{N}(\boldsymbol{y}) = \mathscr{\phi}(\boldsymbol{W}\cdot\boldsymbol{y} + \boldsymbol{b})$$. 
The goal of training an ANN is to tune the values of $\boldsymbol{W}^n$ and $\boldsymbol{b}^n$ for all layers, $n$, so that the desired objective function is optimized. 
Two tasks are carried out in each step of an optimization loop. 
In the forward pass, inputs are fed to the ANN in the IL to obtain the non-linear map $\mathscr{N}^{nL+1}$ at the OL for the purpose of calculating the loss function that is defined w.r.t. the objective of the formulation. 
In the most important back-propagation step, $\boldsymbol{W}^n$ and $\boldsymbol{b}^n$ are recalculated/ updated for all layers $n$ based on the sensitivity of these parameters w.r.t. the errors obtained during the forward pass, i.e., by minimizing the loss function in the process. 
After the training is complete, during prediction, the non-linear mapping from input to the outputs is achieved very quickly, often instantaneously, using minimal computational resources.

\subsection{Input features}

%Appropriate strategies augment a DL model's capability to efficiently learn the complex characteristics underlying the data. 
Carefully chosen input features can significantly improve a model's learning capability during training. 
Functionally, the SGS models provide dissipation to the grid-scale flow field. 
From the DNS of isotropic turbulence, it has been shown that the energy dissipation is concentrated in eddies and convergence zones \citep{hunt1988eddies}. 
In eddies, vorticity dominates the irrotational strain. 
On the other hand, irrotational strain is dominant in convergence zones. 
Therefore, researchers have traditionally utilized flow features such as the velocity gradient tensor ($\nabla \boldsymbol{\tilde{u}}$), the strain rate tensor ($\boldsymbol{\tilde{S}}$), and the rotation rate tensor ($\boldsymbol{\tilde{R}}$), for the formulation of SGS models. 
Additionally, for the DL models to be invariant to Galilean transformation, both the input and output features should be Galilean invariant (invariant under the effect of translation of the coordinate system). 
Researchers have also emphasized the fact that the rotational and reflectional invariances of the feature variables are satisfied \citep{durbin2011statistical, wu2018physics}.

Considering these constraints, several model training methodologies may be envisaged. 
The following list shows a few of the possible strategies, some of which were attempted during the course of this work. 
\begin{itemize}
\item SGS models may be trained to directly approximate $\tilde{\boldsymbol{\tau}}\approx \tilde{\boldsymbol{\tau}}(features)$. 
\citet{gamahara2017searching} trained a separate NN for each component of $\tilde{\boldsymbol{\tau}}$. 
\citet{park2021toward} only used one NN to predict the six independent components of $\tilde{\boldsymbol{\tau}}$. 
These works used the components of either $\nabla\tilde{\boldsymbol{u}}$, $\tilde{\boldsymbol{S}}$, or $\tilde{\boldsymbol{R}}$ (the rotation-rate tensor) as input features. 
Although, the tensors themselves satisfy Galilean invariance, the developed models by these authors do not satisfy this property because of their choice of model architecture. 
\item As in eddy-viscosity type SGS models, the SGS eddy viscosity $\nu_{SGS}$ may be extracted from the DNS flow field. 
However, extraction of $\nu_{SGS}$ involves a least-squares fitting, and as a consequence, the extraction process itself incurs significant errors \citep{matai2019large}. 
\item The SGS stresses may be predicted so that $\tilde{\boldsymbol{\tau}} \approx \tilde{\boldsymbol{\tau}}(\tilde{\boldsymbol{S}}, \tilde{\boldsymbol{R}}, \boldsymbol{I})$ following the effective eddy-viscosity hypothesis formulated by \citet{pope1975more}. 
\end{itemize}

The last methodology was pursued in detail and is the subject of this paper. 
Both Galilean, rotational and reflectional invariances are intrinsically inherited in this analytical framework. 
The general expression for $\tilde{\boldsymbol{\tau}} \approx \tilde{\boldsymbol{\tau}}(\tilde{\boldsymbol{S}}, \tilde{\boldsymbol{R}}, \boldsymbol{I})$ is an infinite tensor polynomial. 
In the effective eddy-viscosity hypothesis, \citet{pope1975more} utilized the Cayley-Hamilton theorem to expand the general tensor expression for $\tilde{\boldsymbol{\tau}}(\tilde{\boldsymbol{S}}, \tilde{\boldsymbol{R}}, \boldsymbol{I})$. 
Ten isotropic basis tensors were required to complete the integrity basis for an incompressible fluid. 
The coefficients of the integrity basis tensors were expressed as functions of 5 scalar invariants of $\boldsymbol{\tilde{S}}$ and $\boldsymbol{\tilde{R}}$. 
Later, \citet{lund1993parameterization} simplified the hypothesis of \citet{pope1975more} in the context of SGS modeling. 
In their formulation, the expansion for the SGS stress tensor is as follows: 

\begin{eqnarray}
\label{Pope}
\tilde{\boldsymbol{\tau}} = \sum_{k=1}^{5}g^{(k)}(\lambda_1, \lambda_2, \lambda_3, \lambda_4, \lambda_5, \lambda_6) \boldsymbol{T}^{(k)}(\tilde{\boldsymbol{S}}, \tilde{\boldsymbol{R}}, \boldsymbol{I})
\end{eqnarray}

In this expression, the 6 scalar invariants are expressed as:

\begin{equation}
\label{inv}
  \begin{split}
    &  \lambda_1 = Tr(\boldsymbol{\tilde{S}^2}),  \hspace{8mm}   \lambda_2 = Tr(\boldsymbol{\tilde{R}^2}),   \hspace{8mm}  \lambda_3 = Tr(\boldsymbol{\tilde{S}^3}), \\ &  \lambda_4 = Tr(\boldsymbol{\tilde{R}^2\tilde{S}}), \hspace{4mm}   \lambda_5 = Tr(\boldsymbol{\tilde{R}^2\tilde{S}^2}),   \hspace{4mm}   \lambda_6 = Tr(\boldsymbol{\tilde{S}^2\tilde{R}^2\tilde{S}\tilde{R}}) 
  \end{split}
\end{equation}

The following are the expressions for the 5 integrity basis tensors:

\begin{equation}
\label{int-base}
  \begin{split}
    &  \boldsymbol{T}^{(1)} = \boldsymbol{\tilde{S}},  \hspace{8mm}   \boldsymbol{T}^{(2)} = \boldsymbol{\tilde{S}\tilde{R}}-\boldsymbol{\tilde{R}\tilde{S}},   \hspace{8mm}  \boldsymbol{T}^{(3)} = \boldsymbol{\tilde{S}^2}-\frac{1}{3}Tr(\boldsymbol{\tilde{S}^2})\boldsymbol{I}, \\ &  \boldsymbol{T}^{(4)} = \boldsymbol{\tilde{R}^2}-\frac{1}{3}Tr(\boldsymbol{\tilde{R}^2})\boldsymbol{I}, \hspace{8mm}   \boldsymbol{T}^{(5)} = \boldsymbol{\tilde{R}\tilde{S}^2}-\boldsymbol{\tilde{S}^2\tilde{R}} 
  \end{split}
\end{equation}

In the present work, the input features are the 6 scalar invariants, and the 5 basis tensors in Eqs. \ref{inv} and \ref{int-base}, respectively. 
Two different strategies have been used, one that embeds all the invariance properties ($A1$), and the other specifically usable in a Galilean invariant reference frame in wall-bounded turbulence ($A2$). 
In approach $A1$, the neural network architecture is such as to emulate the analytical model form given in Eq. \ref{Pope}. 
Because, the inputs to the models are the scalar invariants in Eq. \ref{inv} and the outputs are also scalars, the scalar coefficients, $g^{(k)}$ in Eq. \ref{Pope}, approach $A1$ yields models that are invariant to Galilean, rotational and reflectional transformations. 
In approach $A2$, the invariance properties are relaxed by using a simpler network architecture as in \cite{park2021toward}, for example. 
As the input features are derivatives of the velocity components, this family of models satisfies only the Galilean invariance property. 
These models do not satisfy rotational and reflectional invariances. 
The inputs to the NNs in approach $A2$ are the 36 ($\tilde{x}_i$ with $i=1, 2, ...,36$) input features (each of the 6 independent components of the 5 symmetric basis tensors, i.e., 30 input features and the 6 scalar invariants) that may be constructed from Eqs. \ref{inv} and \ref{int-base}. 
As our intention is to develop models for incompressible flows, the outputs of the NNs are the deviatoric part of $\tilde{\boldsymbol{\tau}}$, as it is customary to absorb the isotropic part of the stress in a modified pressure term. 
Consequently, only the deviatoric components of tensors in Eq. \ref{int-base} are used for both sets of models. 
Finally, the NN models generated in this work are local in nature, i.e., being fed the input features at a spatial location at a given time, the NNs predict the SGS stresses at the same location, at the same time.

\subsection{Network architecture}

\begin{figure}[!ht]
    \begin{center}
        \includegraphics[trim=0 0.1cm 0.1cm 0,clip,width=0.85\textwidth]{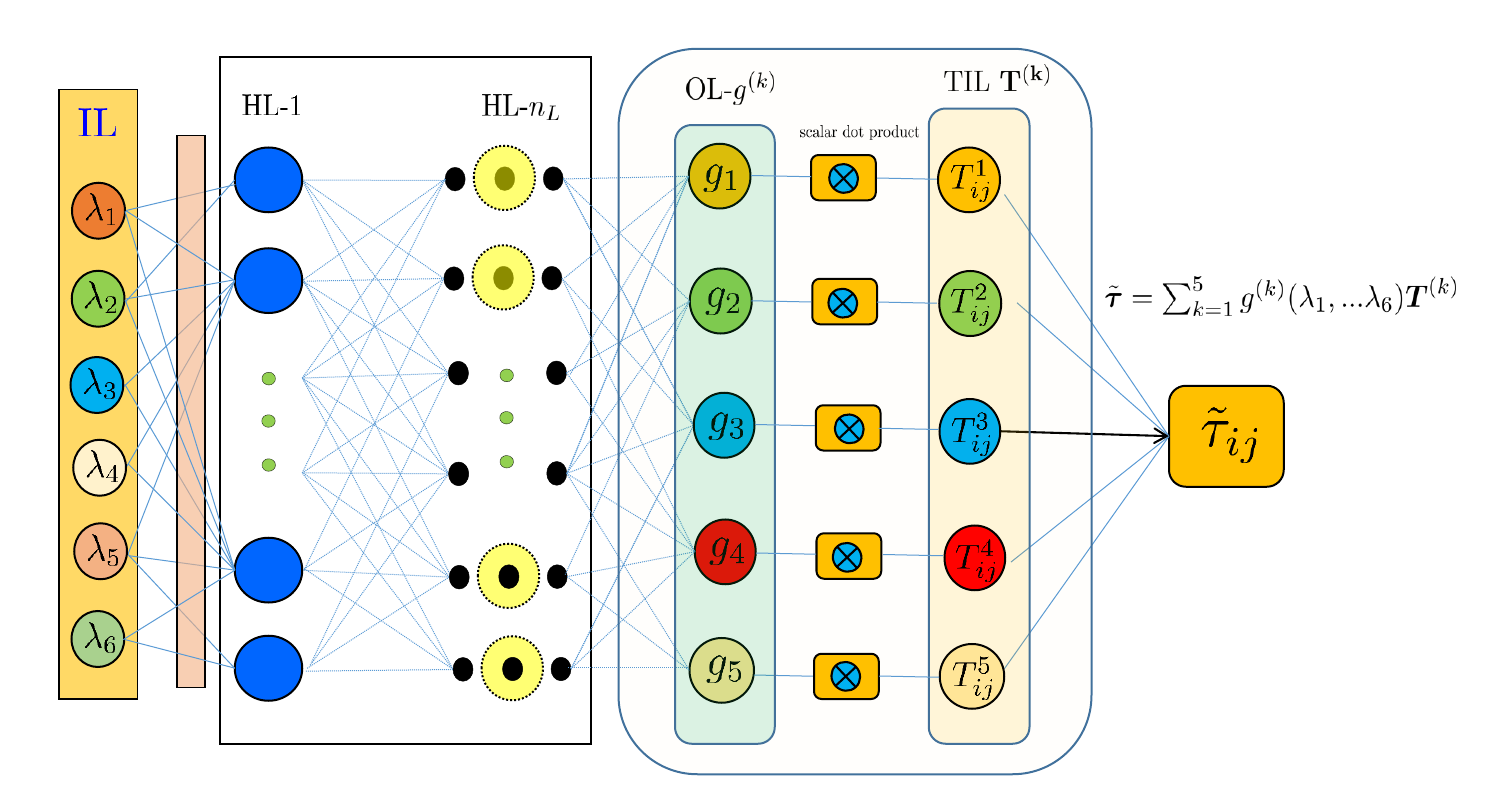}
        \caption{Representative architecture of the NNs for models $A1$. 
        }
        \label{fa1}
    \end{center}
\end{figure} 

The architecture of the NNs in approach $A1$ is that of a tensor basis neural network (TBNN) that was first developed for the purpose of Reynolds stress modeling for Reynolds-averaged Navier-Stokes (RANS) equations by \citet{ling2016reynolds}. 
The architecture of a TBNN is shown in Fig. \ref{fa1}. 
In TBNNs, two separate input layers are used to input the 6 scalar invariants (Eq. \ref{inv}), and the 5 basis tensors (Eq. \ref{int-base}). 
The first input layer takes in as inputs, the scalar invariants to output the 5 scalar coefficients to the basis tensors in the final hidden layer, named $OL-g^{(k)}$ in Fig. \ref{fa1}. 
The tensors are also input using another input layer, called the tensor input layer (TIL), later in the network. 
Finally, in the output layer, the 6 independent components of $\tilde{\boldsymbol{\tau}}$ are the outputs obtained by summing the basis tensors multiplied by the output coefficients in the layer $OL-g^{(k)}$. 
Several hidden layers are used between the invariant input layer and the layer $OL-g^{(k)}$ to extract the complex features in the underlying data. 
In this manner, the TBNN is able to incorporate all the necessary invariant properties by incorporating the analytical model form in Eq. \ref{Pope} in a DL framework. 
It is able to include the contributions of the invariants as well as the basis tensors in the model outputs. 

\begin{figure}[!ht]
    \begin{center}
        \includegraphics[trim=0 0.1cm 0.1cm 0,clip,width=0.85\textwidth]{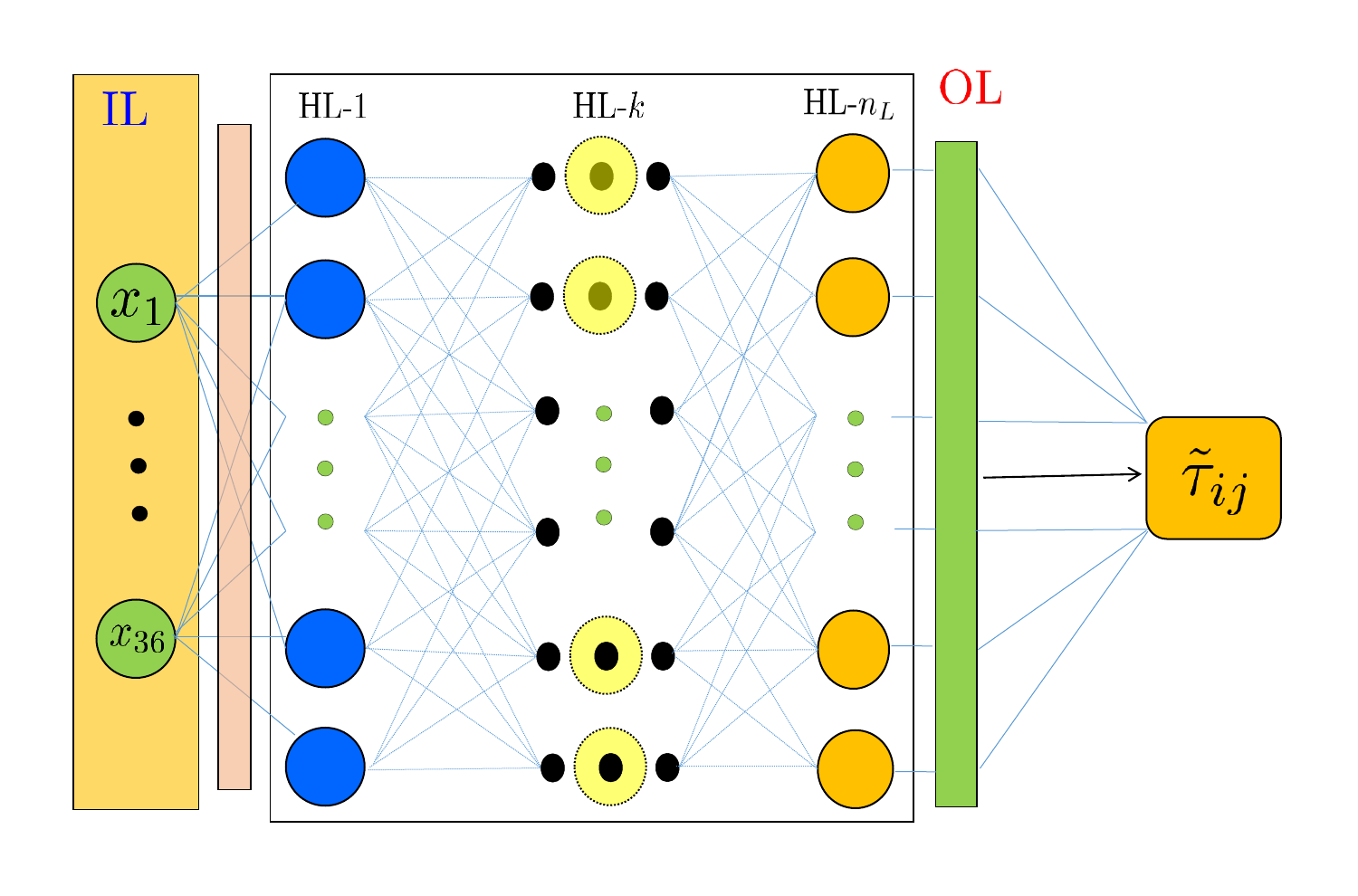}
        \caption{Representative architecture of the NNs for models $A2$.
        }
        \label{fa2}
    \end{center}
\end{figure} 

The architecture of the NNs used in approach $A2$ is much simpler. 
The base architecture is that of a standard feed-forward fully connected NN shown in Fig. \ref{fa2}. 
At the input layer (IL), 36 input features are fed. 
Between IL and the output layer (OL), several hidden layers are used for superior feature extraction. 
The outputs are again the 6 independent components of the SGS tensor $\tilde{\boldsymbol{\tau}}$. 
As the inputs include the components of tensors, this architecture is unable to incorporate the rotational and reflectional invariance properties. 
However, as previously mentioned, the components of the tensors are basically spatial derivatives of the components of the resolved velocity vector, and therefore, are invariant to Galilean transformation. 
As the outputs are also Galilean invariant, these models are applicable in a reference frame fixed at the wall/ translating with a constant velocity. 

There are three hyper-parameters related to the architecture of the NNs used in $A1$ and $A2$. 
One of these is the number of hidden layers ($n_L$). 
Higher $n_L$ generally indicates a more complex network capable of learning more intricate features from data. 
In the present work, we have trained models with $n_L=2,3$ and $4$ (see Tabs. \ref{tA1} and \ref{tA2}). 
The second hyper-parameter is given by the ratio of the number of neurons in each layer and the number of input features in the input layer (named $\beta$). 
In our training methodology, we have varied $\beta$ so that two networks from $A1$ and $A2$ with the same number of hidden layers, $n_L$, have the same number of neurons in each hidden layer. 
In this way, fair comparisons can be made between the NNs from the two families with same $n_L$, as these would have approximately similar number of trainable parameters. 
The third hyper-parameter is the activation function used to facilitate learning of non-linear characteristics in the data (the activation function in the $n^{th}$ layer of a NN is indicated by $\phi^n$ in Eq. \ref{nn}). 
In the present study, we have used a chosen activation function in all the layers of a NN. 
It must be noted however, that this might not yield the most accurate NN. 
In total, we have tested four activation functions, Rectified Linear Unit ($ReLU$), Leaky Rectified Linear Unit (Leaky $ReLU$), the hyperbolic tangent ($\tanh$), and the sigmoid ($\sigma$) activation functions (see Tabs. \ref{tA1} and \ref{tA2}). 
The slope of the function for input $x < 0$ for the Leaky $ReLU$ activation was taken as $\alpha_{LR}=0.05$.

\subsection{Loss function \& fixed hyper-parameters}

Several loss functions were attempted in the course of this work. 
The following loss function, a variant of the mean squared error ($m.s.e.$) was used: 

\begin{equation}
\label{loss}
%\mathscr{L} = (\tilde{\tau}_{11}^p - \tilde{\tau}_{11}^{FF})^2 + (\tilde{\tau}_{22}^p - \tilde{\tau}_{22}^{FF})^2 + (\tilde{\tau}_{33}^p - \tilde{\tau}_{33}^{FF})^2 + (\tilde{\tau}_{12}^p - \tilde{\tau}_{12}^{FF})^2 + (\tilde{\tau}_{23}^p - \tilde{\tau}_{23}^{FF})^2 + \alpha_\mathscr{L}\boldsymbol{W}^0\cdot\boldsymbol{W}^0  
\mathscr{L} = \frac{1}{N_b}\sum_{N_b}(\boldsymbol{\tilde{\tau}}^p - \boldsymbol{\tilde{\tau}}^{FF})\cdot(\boldsymbol{\tilde{\tau}}^p - \boldsymbol{\tilde{\tau}}^{FF}) + \alpha_\mathscr{L}\sum_{n=1} ^{nL}|\boldsymbol{W}^n|^2  
\end{equation}

In this expression, $N_b$ is the batch size, $\boldsymbol{\tilde{\tau}}^p$ is the SGS stress tensor predicted by the NN and $\boldsymbol{\tilde{\tau}}^{FF}$ is the true stress tensor obtained from the FF. 
As the NN predicts the 6 components of the SGS stresses, only these 6 components are considered while computing the dot product. 
The second term on the RHS of Eq. \ref{loss} corresponds to the magnitude-squared weights of the NN (see Eq. \ref{nn}). 
Here, $\alpha_\mathscr{L}$ is a tuning parameter. 
We used $\alpha_\mathscr{L}=0.005$ following the suggestion by \citet{park2021toward}. 
Inclusion of this term yielded slightly more accurate statistics for the NN model predictions compared to the standard $m.s.e.$ loss function.

The NNs are trained by minimizing the loss function using the Adam optimization algorithm \citep{kingma2014adam}. 
During training, input data is fed to the NN to compute the loss function. 
The weights and biases of a network are updated in a backpropagation step by computing the sensitivity of the loss function to the trainable parameters of the NN several times in a training epoch. 
An epoch represents one pass of the complete training data set through the NN in a forward pass. 
In our training procedure, 500 epochs were used to train each network. 
The loss function saturated to a minima well within the completion of the 500 training epochs (see Figs. \ref{feA1} and \ref{feA2}). 
The trainable parameters are updated several times within a training epoch. 
The number of times the trainable parameters in a NN are updated in a training epoch is dictated by the batch size. 
Following \citet{kingma2014adam}, the batch size was set to $N_b=128$. 
The learning rate dictates the rate of change of the trainable parameters in a NN based on the computed gradients of the loss function w.r.t. these parameters. 
Considering that a large number of epochs were used for training, we used a learning rate of 0.001 so that abrupt large changes in the trainable parameters of the NN could be avoided, and these smoothly converged to the desired optimum. 
To avoid over-fitting of the NNs to the training data, a regularization technique called dropout is used. 
The dropout indicates the number of neurons that are randomly switched off in each layer of a NN during training. 
We used a dropout value of 0.1 which indicates that $10\%$ of the neurons in each layer are randomly switched off during a forward pass in training. 
The value of the loss function for the validation data is checked at the end of each training epoch, and the weights and biases of the NN are saved in case the obtained validation loss is lower than the minimum validation loss recorded until that point in training. 
Finally, training of the NNs is performed utilizing the Keras DL library with tensorflow \citep{abadi2016tensorflow} as its backend.

\subsection{Data compilation \& scaling}

Data from the 120 time snapshots of the whole spatial domain stored from each DNS (see Sec. \ref{sec2}) was filtered and the desired input features and output SGS stresses were generated. 
Time instants were randomly chosen without replacement, and stored for generating training, validation and test datasets for all the models. 
%Data from the same time snapshots were always used for training, validation, and testing all the models. 
80\% of the time snapshots were used for training, 15\% for validation, and the rest 5\% for testing. 
%In terms of snapshots, 
Out of 120 time snapshots, data from 96 and 18 time snapshots were used for generating the training and validation data sets, respectively; the remaining 6 time snapshots were used for testing all models. 

At the beginning of training of a NN model, data tuples from the time snapshots segregated for training and validation were further sampled to generate the final datasets for training and validating a model. 
A data tuple consists of the input features and output SGS stresses generated at a given location at a given time. 
On average, only one data tuple is chosen to be in the final training and validation datasets, picked randomly without replacement, out of every 32 available data tuples. 
The final training and validation datasets for all the models contained 750384 and 140697 data tuples, respectively. 
Instead of using all the data tuples, this sampling procedure was adopted to avoid feeding the NNs with spatially correlated data. 
In this way, the NNs are able to learn complex functional relations between the inputs and the outputs. 
This quickens the feature extraction procedure making the training process efficient, and the resultant models, robust. 
Sampling of this kind was not necessary during testing the model performance. 
There were 1500768, 2668032, and 4496472 data tuples in the three test datasets tabulated in Tab. \ref{t2}, respectively.  

Another important parameter for training DL-SGS models is the scaling of the datasets. 
Several possibilities may be envisaged, including the normalization, standardization, and scaling based on fluid dynamic variables. 
\citet{park2021toward} used a scaling based on fluid dynamic variables, channel centreline velocity and channel half height, $\delta$. 
The trained models performed poorly for a higher-$Re$ flow with a NN trained at lower $Re$. 
During the course of the present work, several scalings were attempted. 
The normalization was attempted by extracting the maximum and minimum from the whole training dataset. 
In another normalization scaling strategy, data normalization was performed based on distance from the wall. 
However, both of these scaling strategies failed to yield good model performance. 
The standardization also did not achieve good model performance. 
Eventually, a fluid dynamic scaling, with friction velocity ($u^*$) and the viscous length scale ($\nu/u^*$) as the velocity and length scales yielded successful models. 
This scaling has the advantage that it is applicable in any wall-bounded flow where the friction velocity may be calculated instantaneously, or in a priori simulations that may used for scaling the data.

\begin{table}
\scriptsize
\begin{tabular*}{\linewidth}{@{\extracolsep{\fill}} c c c c c c c c }
\hline
Activation ($\phi$) & $\beta$ & $n_L$ & $L_v$ & $L_t$ & $\varepsilon_v$ & $\varepsilon_t$ & $t_e$\\
%\cline{2-5}
% & \multicolumn{2}{c|}{Value} & \multicolumn{2}{c|}{Value} & \\
%\cline{2-5}
\hline
                           &  6 & 2 & 0.5714 & 1.1398 & 0.5165 & 1.0848 & 4.82 \\
                           &  6 & 3 & 0.5585 & 0.5616 & 0.5035 & 0.5066 & 5.0 \\
                           &  6 & 4 & 0.5548 & 0.5558 & 0.5029 & 0.5039 & 5.96\\
                           & 30 & 2 & 0.5746 & 0.5729 & 0.516 & 0.5143 & 7.01 \\
Leaky $ReLU$               & 30 & 3 & 0.5635 & 0.5752 & 0.5051 & 0.5168 & 9.01 \\
                           & 30 & 4 & 0.5643 & 0.6406 & 0.5038 & 0.5801 & 11.0 \\
                           & 60 & 2 & 0.5805 & 0.8465 & 0.5227 & 0.7886 & 9.79 \\
                           & 60 & 3 & 0.5682 & 0.5884 & 0.5094 & 0.5295 & 13.99 \\
                           & 60 & 4 & 0.5642 & 0.7707 & 0.5056 & 0.7121 & 19.07 \\
\hline
                             & 6 & 2 & 0.5432 & 0.5441 & 0.4912 & 0.4921 & 4.47 \\
                             & 6 & 3 & 0.5379 & 0.545 & 0.4846 & 0.4916 & 4.97 \\
                             & 6 & 4 & 0.5507 & 0.5463 & 0.4984 & 0.4941 & 5.03 \\
                             & 30 & 2 & 0.5455 & 0.5431 & 0.4906 & 0.4882 & 6.88 \\
$ReLU$                       & 30 & 3 & 0.5526 & 0.5453 & 0.4931 & 0.4858 & 8.46 \\
                             & 30 & 4 & 0.5579 & 0.5506 & 0.4987 & 0.4914 & 10.07 \\
                             & 60 & 2 & 0.5574 & 0.5481 & 0.4964 & 0.4871 & 9.08 \\
                             & 60 & 3 & 0.5617 & 0.5467 & 0.5038 & 0.4888 & 13.11 \\
                             & 60 & 4 & 0.5475 & 0.5483 & 0.4892 & 0.4899 & 17.85 \\
\hline
                             & 6 & 2 & 0.5694 & 0.5627 & 0.5173 & 0.5106 & 4.81 \\
                             & 6 & 3 & 0.5572 & 0.5635 & 0.4946 & 0.501 & 5.0 \\
                             & 6 & 4 & 0.5541 & 0.5503 & 0.5058 & 0.5021 & 5.1 \\
                             & 30 & 2 & 0.5547 & 0.5594 & 0.4982 & 0.5029 & 6.68 \\
$\tanh$                       & 30 & 3 & 0.5592 & 0.5535 & 0.5013 & 0.4956 & 8.14 \\
                             & 30 & 4 & 0.5607 & 0.5533 & 0.5034 & 0.496 & 10.05 \\
                             & 60 & 2 & 0.5673 & 0.5619 & 0.5054 & 0.5 & 8.91 \\
                             & 60 & 3 & 0.5582 & 0.5604 & 0.4979 & 0.5001 & 13.02 \\
                             & 60 & 4 & 0.5683 & 0.5673 & 0.5078 & 0.5068 & 17.53 \\
\hline
                             & 6 & 2 & 0.55 & 0.545 & 0.5005 & 0.4955 & 4.98 \\
                             & 6 & 3 & 0.5428 & 0.549 & 0.4932 & 0.4994 & 5.0 \\
                             & 6 & 4 & 0.5509 & 0.55 & 0.5002 & 0.4993 & 6.14 \\
                             & 30 & 2 & 0.5536 & 0.5511 & 0.4948 & 0.4922 & 7.05 \\
$\sigma$                    & 30 & 3 & 0.5419 & 0.5476 & 0.4858 & 0.4915 & 9.0 \\
                             & 30 & 4 & 0.5534 & 0.5515 & 0.4957 & 0.4938 & 11.0 \\
                             & 60 & 2 & 0.5572 & 0.5502 & 0.4979 & 0.491 & 9.0 \\
                             & 60 & 3 & 0.5544 & 0.5532 & 0.4941 & 0.4929 & 13.07 \\
                             & 60 & 4 & 0.564 & 0.552 & 0.5056 & 0.4936 & 17.92 \\                             
\hline
\end{tabular*}
\caption{\label{tA1}All models trained using NN architecture $A1$ shown in Fig. \ref{fa1}: $\beta$ is the ratio of the number of neurons in each hidden layer and the number of scalar input features (6 for the $A1$ models), and $n_L$ is the number of hidden layers. $L_v$ and $L_t$ are the validation and training losses, $\varepsilon_v$ and $\varepsilon_t$ are the validation $m.s.e.$ and training $m.s.e.$, respectively; Average time (in seconds) taken in each epoch of model training is indicated by $t_e$.}
\end{table}

\subsection{Hyper-parameter tuning}

In this section, we present the results of the study undertaken to tune the hyper-parameters $\beta$, $n_L$, and the activation function to yield optimal model performance. 
Table~\ref{tA1} tabulates all the 36 networks trained using approach $A1$. 
Apart from the three hyper-parameters, the recorded training and validation losses ($L_t$ and $L_v$ respectively), the training and validation $m.s.e$ ($\varepsilon_t$ and $\varepsilon_v$ respectively) are listed in this table for the final model weights and biases. 
The average time taken in each epoch ($n_e$) of model training ($t_e$) over the 500--epoch training duration is also listed. 
Clearly, the more complex models (with more neurons and layers) require longer to train. 
Apart from the Leaky $ReLU$ activation, the difference between the $L_v$ and $L_t$ for the trained models are small indicating that the models are not over-fitted to the training data. 
There is some difference in recorded errors for the models trained with different activation functions. 
However, the differences are not large, and may be insignificant.

\begin{figure}[!ht]
    \begin{center}
        \includegraphics[trim=0 0cm 0cm 0,clip,width=0.99\textwidth]{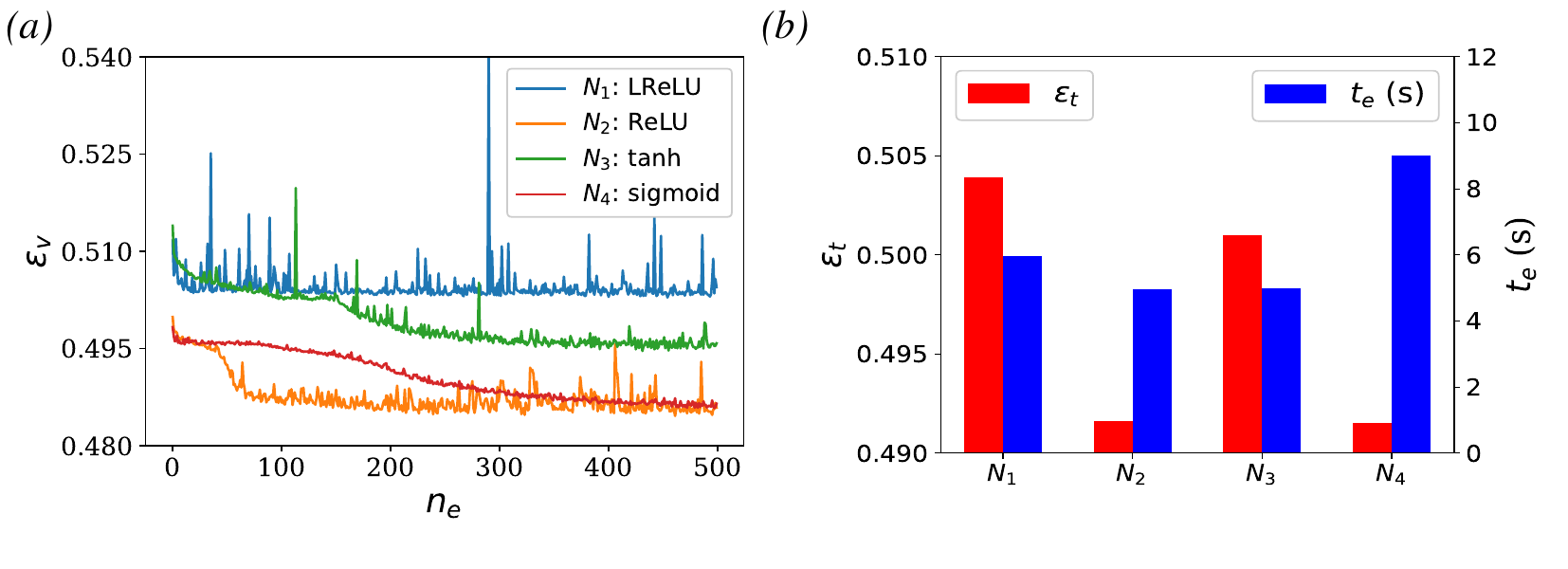}
        \caption{($a$) Validation m.s.e. ($\varepsilon_v$) plotted against epochs ($n_e$) for the four chosen models corresponding to each of the activation functions used in the architecture of $A1$ models as listed in Tab. \ref{tA1}. 
        ($b$) Comparisons of training m.s.e. ($\varepsilon_t$) and time required in each epoch of model training ($t_e$) in seconds for these four chosen models. 
        }
        \label{feA1}
    \end{center}
\end{figure} 

Based on this table, we have chosen the model that yielded the minimum $L_v$ among the 9 models trained with each activation function. 
These four models are named $N_1$, $N_2$, $N_3$, and $N_4$ corresponding to the Leaky $ReLU$, $ReLU$, $\tanh$, and $\sigma$ activation, respectively. 
The ($\beta, n_L$) combinations for these four models are (6, 4), (6, 4), (6, 3), and (30, 3), respectively. 
These values indicate that NNs with more hidden layers were more efficient at fitting the training data compared to NNs with more neurons in each hidden layer. 
Figure~\ref{feA1}($a$) shows $\varepsilon_v$ plotted against $n_e$ over the whole training duration for all the four models. 
$\varepsilon_v$ is lower for the $ReLU$ and $\sigma$ activations, and largest for the Leaky $ReLU$ activation. 
Models trained with the Leaky $ReLU$ activation shows large wiggles in $\varepsilon_v$ during training, indicating that this activation may not be appropriate for the $A1$ models. 
Figure~\ref{feA1}($b$) shows bar charts for $\epsilon_t$ and $t_e$ for the four chosen models for a clearer comparison. 
The values are extracted from Tab. \ref{tA1}. 
As the training dataset consists of significantly more data compared to the validation dataset, and the difference between $\varepsilon_t$ and $\varepsilon_v$ are small, $\varepsilon_t$ reflects a model's performance over a large volume of data. 
The bar charts also show that models $N_2$ and $N_4$ provide more accurate predictions. 
However, significantly longer time was required to train model $N_4$ because this model had larger number of trainable parameters compared to the other three models.

\begin{table}
\scriptsize
\begin{tabular*}{\linewidth}{@{\extracolsep{\fill}} c c c c c c c c }
\hline
Activation ($\phi$) & $\beta$ & $n_L$ & $L_v$ & $L_t$ & $\varepsilon_v$ & $\varepsilon_t$ & $t_e$\\
%\cline{2-5}
% & \multicolumn{2}{c|}{Value} & \multicolumn{2}{c|}{Value} & \\
%\cline{2-5}
\hline
                           &  1 & 2 & 0.6048 & 0.6073 & 0.425 & 0.4274 & 4.79 \\
                           &  1 & 3 & 0.601 & 0.6053 & 0.4246 & 0.4289 & 5.0 \\
                           &  1 & 4 & 0.6076 & 0.6066 & 0.4301 & 0.4291 & 5.2 \\
                           &  5 & 2 & 0.7146 & 0.72 & 0.4155 & 0.4208 & 7.0 \\
Leaky $ReLU$               &  5 & 3 & 0.7224 & 0.7188 & 0.4229 & 0.4193 & 8.98 \\
                           &  5 & 4 & 0.7161 & 0.7134 & 0.4178 & 0.415 & 10.97 \\
                           & 10 & 2 & 0.7411 & 0.7405 & 0.4165 & 0.4159 & 9.56 \\
                           & 10 & 3 & 0.7424 & 0.7439 & 0.4145 & 0.416 & 14.02 \\
                           & 10 & 4 & 0.7449 & 0.745 & 0.4172 & 0.4173 & 18.91 \\
\hline
                        &  1 & 2 & 0.6069 & 0.6115 & 0.4212 & 0.4258 & 4.76 \\
                        &  1 & 3 & 0.6 & 0.6013 & 0.4245 & 0.4257 & 4.97 \\
                        &  1 & 4 & 0.6034 & 0.6111 & 0.4177 & 0.4254 & 5.0 \\
                        &  5 & 2 & 0.7214 & 0.7173 & 0.42 & 0.4159 & 6.35 \\
$ReLU$                  &  5 & 3 & 0.7241 & 0.7135 & 0.4264 & 0.4157 & 8.2 \\
                        &  5 & 4 & 0.725 & 0.716 & 0.4299 & 0.4209 & 10.0 \\
                        & 10 & 2 & 0.7531 & 0.7458 & 0.423 & 0.4157 & 9.0 \\
                        & 10 & 3 & 0.7532 & 0.7425 & 0.4246 & 0.4139 & 13.15 \\
                        & 10 & 4 & 0.7511 & 0.7474 & 0.425 & 0.4213 & 17.87 \\
\hline
                        &  1 & 2 & 0.6166 & 0.6243 & 0.4322 & 0.4399 & 4.43 \\
                        &  1 & 3 & 0.6103 & 0.6208 & 0.4288 & 0.4393 & 5.0 \\
                        &  1 & 4 & 0.6018 & 0.6126 & 0.4325 & 0.4433 & 5.02 \\
                        &  5 & 2 & 0.7268 & 0.7231 & 0.4281 & 0.4244 & 6.57 \\
$\tanh$                  &  5 & 3 & 0.7247 & 0.7235 & 0.4266 & 0.4254 & 8.02 \\
                        &  5 & 4 & 0.7231 & 0.729 & 0.4264 & 0.4323 & 10.0 \\
                        &  10 & 2 & 0.7441 & 0.7543 & 0.417 & 0.4272 & 8.96 \\
                        &  10 & 3 & 0.7606 & 0.7517 & 0.428 & 0.4191 & 13.51 \\
                        &  10 & 4 & 0.7628 & 0.7605 & 0.4373 & 0.435 & 17.73 \\
\hline
                        &  1 & 2 & 0.6142 & 0.6205 & 0.4295 & 0.4357 & 4.7 \\
                        &  1 & 3 & 0.6132 & 0.6124 & 0.4333 & 0.4326 & 4.99 \\
                        &  1 & 4 & 0.6129 & 0.6153 & 0.4308 & 0.4332 & 5.3 \\
                        &  5 & 2 & 0.7268 & 0.714 & 0.429 & 0.4162 & 6.92 \\
$\sigma$               &  5 & 3 & 0.7217 & 0.7148 & 0.4279 & 0.421 & 8.89 \\
                        &  5 & 4 & 0.7326 & 0.7258 & 0.4349 & 0.4281 & 10.97 \\
                        &  10 & 2 & 0.7566 & 0.7419 & 0.4291 & 0.4144 & 9.0 \\
                        &  10 & 3 & 0.7638 & 0.7492 & 0.4359 & 0.4213 & 13.04 \\
                        &  10 & 4 & 0.7604 & 0.7498 & 0.436 & 0.4254 & 18.87 \\                             
\hline
\end{tabular*}
\caption{\label{tA2}All models trained using NN architecture $A2$ shown in Fig. \ref{fa2}: $\beta$ is the ratio of the number of neurons in each hidden layer and the number of scalar input features (6 for the $A1$ models), and $n_L$ is the number of hidden layers. $L_v$ and $L_t$ are the validation and training losses, $\varepsilon_v$ and $\varepsilon_t$ are the validation $m.s.e.$ and training $m.s.e.$, respectively; Average time (in seconds) taken in each epoch of model training is indicated by $t_e$.}
\end{table}

In Tab. \ref{tA2}, the same variables tabulated in Tab. \ref{tA1} for the $A1$ models, are tabulated for the $A2$ models. 
The values of $\beta$ are lower for the models trained in this approach, because, the number of input features are 6 times larger for this class of models. 
Both training and validation $m.s.e.$ are significantly smaller for the $A2$ models compared to the $A1$ models. 
Similar to Tab. \ref{tA1}, the difference between $L_v$ and $L_t$, and between $\varepsilon_t$ and $\varepsilon_v$ are small. 
Hence, it may be assume that the training for the models have been successful, and the models are not overfitted to the training data. 
The difference between the errors recorded for models with same model architecture but using different activation functions are small. 
Additionally, NNs with more trainable parameters have taken significantly longer training duration (see the column for $t_e$). 

\begin{figure}[!ht]
    \begin{center}
        \includegraphics[trim=0 0cm 0cm 0,clip,width=0.99\textwidth]{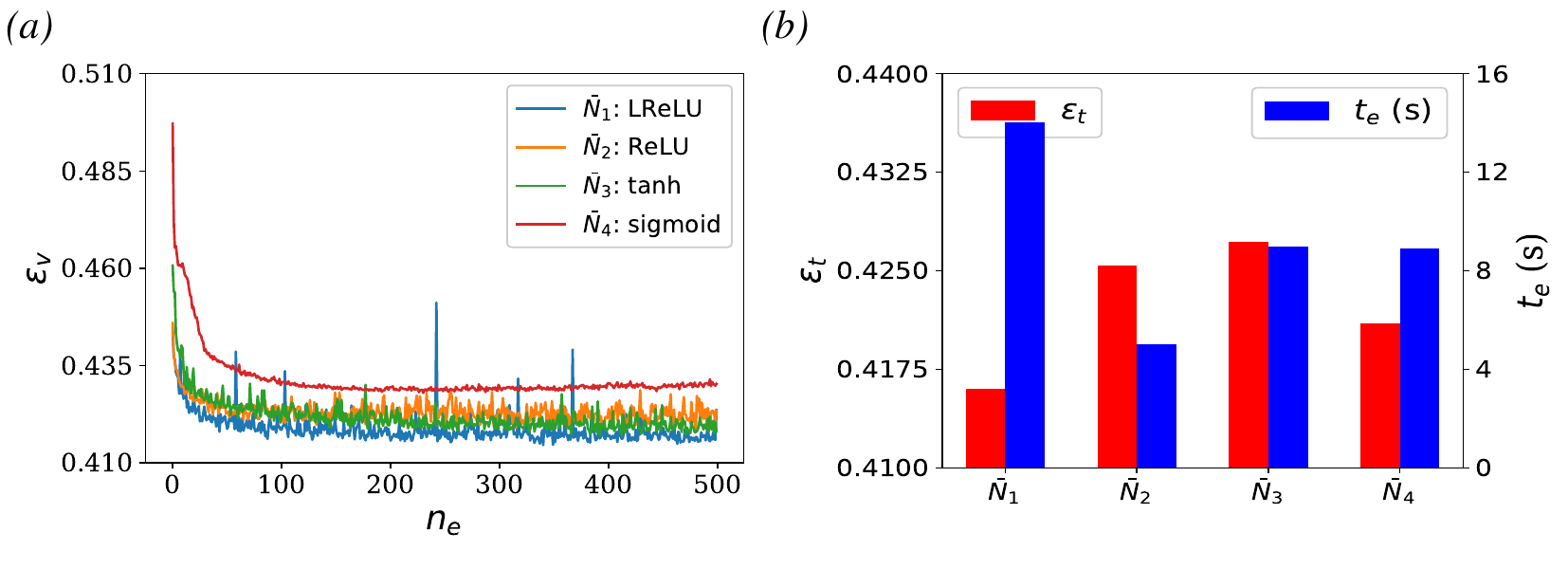}
        \caption{($a$) Validation m.s.e. ($\varepsilon_v$) plotted against epochs ($n_e$) for the four chosen models corresponding to each of the activation functions used in the architecture of $A2$ models as listed in Tab. \ref{tA2}. 
        ($b$) Comparisons of training m.s.e. ($\varepsilon_t$) and time required in each epoch of model training ($t_e$) in seconds for these four chosen models. 
        }
        \label{feA2}
    \end{center}
\end{figure} 

For all the $A2$ models trained, we have also chosen a model corresponding to each activation function that yielded the minimum $L_v$. 
These models are named $\bar{N}_1$, $\bar{N}_2$, $\bar{N}_3$, and $\bar{N}_4$, respectively for the Leaky $ReLU$, $ReLU$, $\tanh$, and $\sigma$ activation functions. 
From Tab. \ref{tA2}, those correspond to ($\beta, n_L$)--combinations, (10, 3), (1, 4), (10, 2), and (5, 3), respectively. 
In Fig. \ref{feA2}($a$), $\varepsilon_v$ is plotted against epochs ($n_e$) for these four models. 
The models smoothly converged to the optimal weights and biases. 
Among these models, $\bar{N}_4$ with $\sigma$ activation recorded higher $\varepsilon_v$ compared to the other three models. 
Figure~\ref{feA2}($b$) shows the bar charts for $\varepsilon_t$ and $t_e$ recorded for these four models. 
Model $\bar{N}_1$ required significantly longer training time because of its significantly more trainable parameters compared to the other three models. 
The bar charts for $\varepsilon_t$ show that despite its complex architecture, it is not significantly more accurate than the other models. 

In the next stage of the work, we compared the performances of the $N_1$--$N_4$ models chosen from all the trained $A1$ family of models. 
For this purpose, we compared the prediction performances of these models on the test data generated on the training grid (dataset $D1$ in Tab. \ref{t2}) for the $Re_\tau \sim 395$ case. 
Among these four models, the model $N_4$ using the $\sigma$ activation with ($\beta, n_L$)=(30, 3) provided the best performance on the test dataset $D1$; this model is hereafter called $M1$. 
The same exercise was performed for the four chosen models from all the $A2$ family of models, $\bar{N}_1$--$\bar{N}_4$. 
Among these four models, $\bar{N}_4$ using the $\sigma$ activation and ($\beta, n_L$)=(5, 3) performed best on the test dataset $D1$. 
We name this model $M2$. 
In the next section, we compare the performances of these two models based on different statistical and fluid dynamic performance metrics. 
This comparison will highlight the advantages and disadvantages of the two approaches used herein to develop the two sets of models, the invariance embedded tensor basis neural networks (model $M1$ represents this approach $A1$), and the more data-driven approach that only embeds Galilean invariance (model $M2$ is representative of approach $A2$).

\section{Model Performance Assessment}\label{sec4} 

Models $M1$ and $M2$ are tested on the features generated on three different datasets, $D1$, $D2$ and $D3$ listed in Tab. \ref{t2}. 
These three datasets are used to test the models' performance on: (i) test data generated on the same grid and $Re$ as the training data (dataset $D1$), (ii) test data generated on a different grid but at the same $Re$ as the training data (dataset $D2$), (iii) test data generated on a similar grid as the training data but at a different $Re$ (dataset $D3$). 
Both statistical and fluid dynamic parameters are used to gauge the performance of the two models. 
The model predictions are compared with the true values of the compared quantities that are obtained from the explicitly filtered DNS data, and denoted as $FF$ in the following discussion.

\subsection{Predictive performance on the same grid and Reynolds number as training}

\begin{figure}[!ht]
    \begin{center}
        \includegraphics[trim=0 0cm 0cm 0,clip,width=0.99\textwidth]{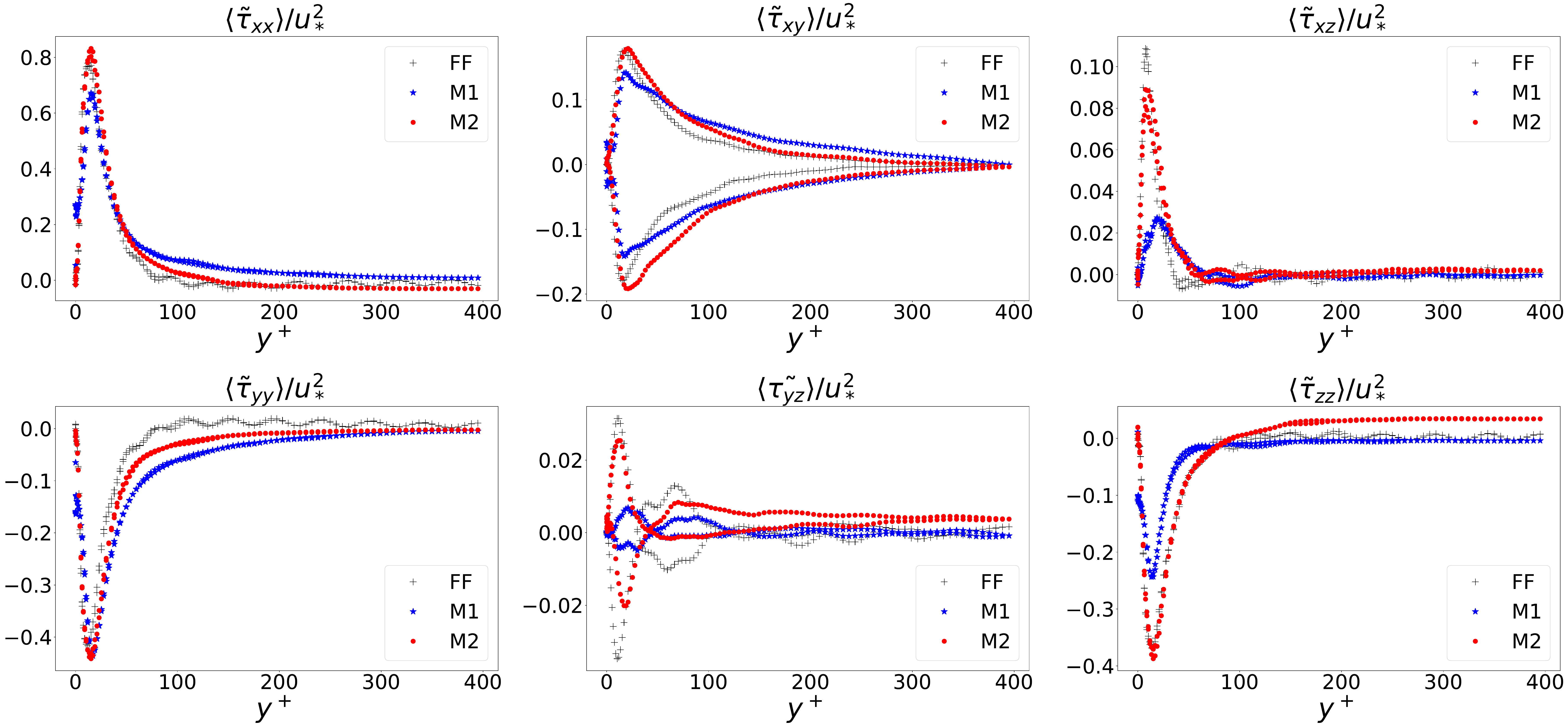}
        \caption{Testing performance of the $M1$ and $M2$ models compared with the filtered field (FF) on the training grid (dataset $D1$ in Tab. \ref{t2}): scaled mean SGS stresses ($\langle \tilde{\tau_{ij}} \rangle$) are shown for the components of the symmetric stress tensor $\tilde{\tau_{ij}}$. 
        %The FF used for the input features to the models $M1$ and $M2$ are from the LES grid $36\times48\times36$ at $Re_\tau \sim 395$ listed in Tab. \ref{t2}. 
        }
        \label{fmtgt}
    \end{center}
\end{figure} 

In Fig. \ref{fmtgt}, the components of the SGS stresses predicted by the two models are averaged in the statistically homogeneous horizontal directions and over time snapshots, are plotted in the inhomogeneous wall-normal direction. 
The SGS stresses are scaled by the friction velocity scale $u^*$, and the wall-normal coordinate is scaled by the viscous wall unit ($y+=yu*/\nu$). 
For comparison, the true SGS stresses are also shown. 
For a LES of a wall-bounded flow, among the components of $\boldsymbol{\tau}$, the off-diagonal stress $\tilde{\tau}_{xy}$ is the most important component along with the $\tilde{\tau}_{xz}$--component. 
Component $\tilde{\tau}_{yz}$ is the least important because of its significantly lower magnitude. 
For all six components, model $M2$ performs notably better than $M1$ at predicting the mean SGS stresses. 
The predictions by $M2$ are in excellent agreement with $FF$ close to the wall for all components. 
The performance only slightly deteriorates away from the wall. 
On the other hand, mean SGS stresses are appreciably underpredicted close to the wall, especially for the cross-component terms by $M1$.

\begin{figure}[!ht]
    \begin{center}
        \includegraphics[trim=0 0cm 0cm 0,clip,width=0.99\textwidth]{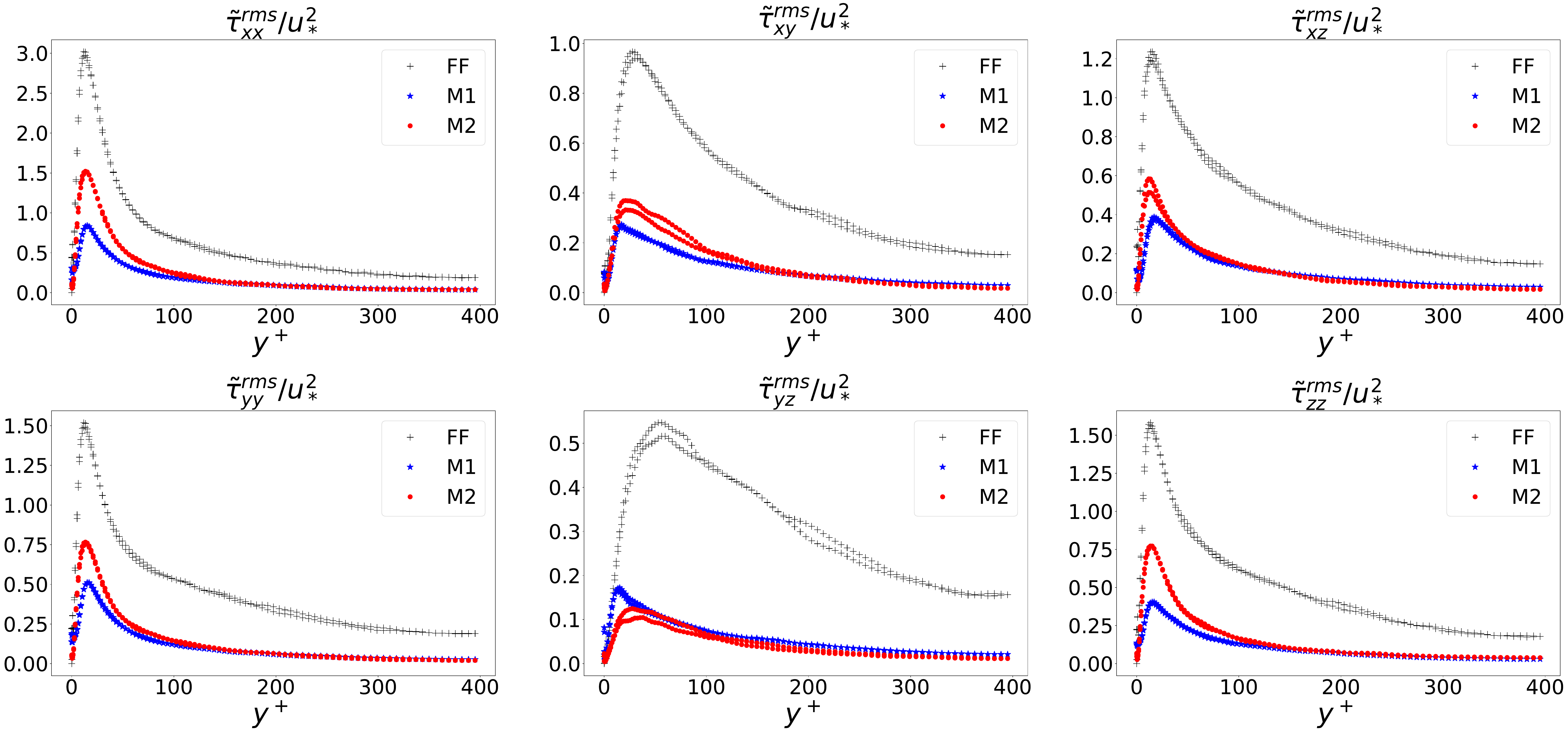}
        \caption{Performance of the $M1$ and $M2$ models compared with the filtered field (FF) for the test dataset generated on the training grid (dataset $D1$ in Tab. \ref{t2}): scaled $r.m.s.$ SGS stresses ($\tilde{\tau}^{rms}_{ij}/u^*$). 
        %The FF used for the input features to the models $M1$ and $M2$ are from the LES grid $36\times48\times36$ at $Re_\tau \sim 395$ listed in Tab. \ref{t2}. 
        }
        \label{frmstgt}
    \end{center}
\end{figure} 

Figure~\ref{frmstgt} compares the wall-normal profiles of the root-mean-squared ($r.m.s.$) SGS stresses predicted by the two models. 
%The FF corresponds to profiles of $r.m.s.$ true SGS stresses. 
This plot reflects the variance of the model predicted SGS stresses and compares those with the variance of the true SGS stresses ($FF$). 
Close to the wall, the variance of the predictions by $M2$ is larger than the $M1$ model for all the components except the $\tilde{\tau}_{yz}$ component. 
Also the characteristics of the profiles near the wall for the two models are similar to the $FF$. 
Differences in the variance of the model predictions reduce away from the wall, where the turbulence is more isotropic in nature. 
This indicates that the $M2$ model is able to learn the anisotropic characteristics of $\boldsymbol{\tilde{\tau}}$ better than $M1$. 
However, the variance provided by both models is relatively low. 
It is noted, however, that although the SGS stresses are predicted by the models, in the LES equations, $\nabla \cdot \boldsymbol{\tilde{\tau}}$ enters the equations (see Eqs. \ref{les1} and \ref{les2}) instead of $\boldsymbol{\tilde{\tau}}$. 
Therefore, the local spatial gradients of the SGS stresses are important, rather than their absolute values. 
This figure is presented from the context of a typical DL model testing perspective. 

\begin{figure}[!ht]
    \begin{center}
        \includegraphics[trim=0 0cm 0cm 0,clip,width=0.99\textwidth]{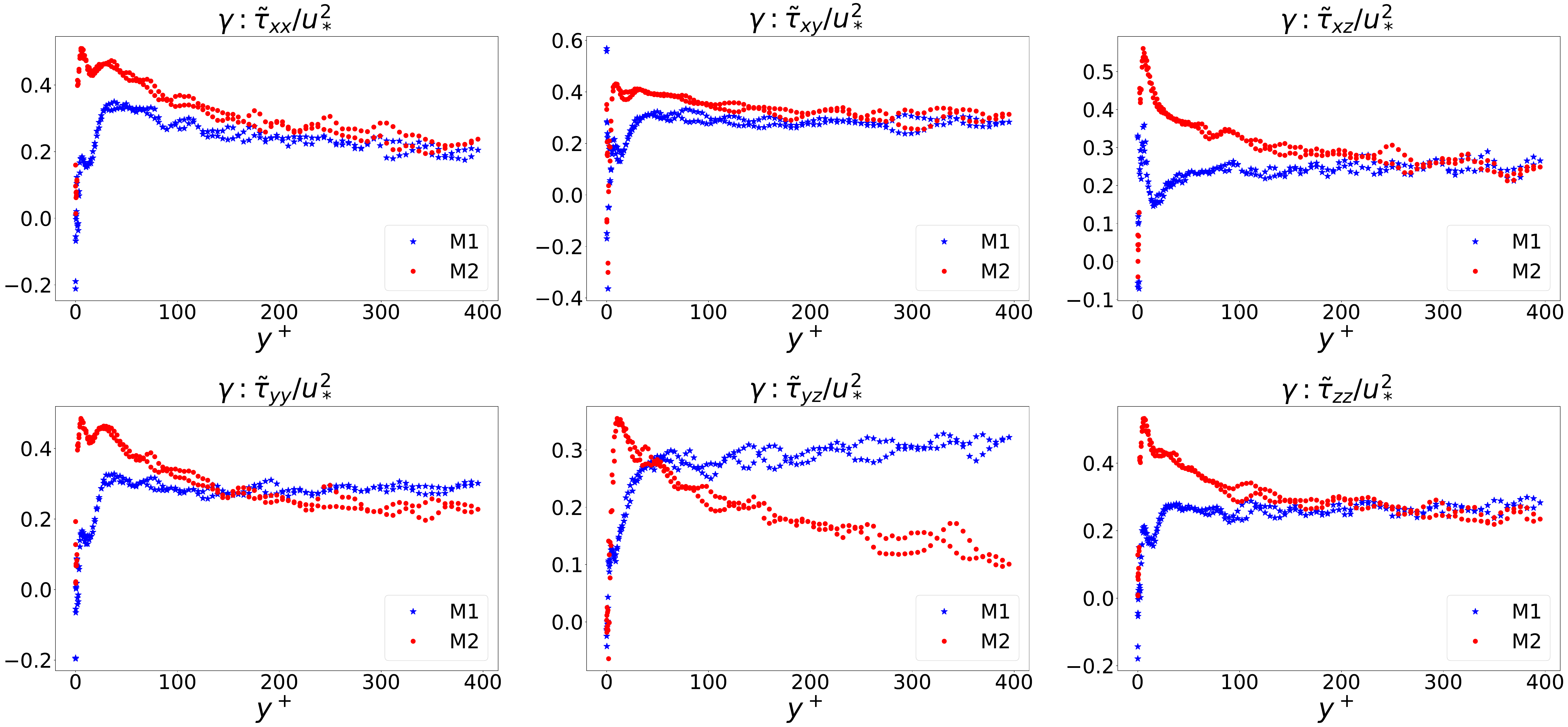}
        \caption{Comparison of the testing performance of the $M1$ and $M2$ models (on dataset $D1$ in Tab. \ref{t2}): correlation coefficient ($\gamma$) computed between the predicted SGS stresses ($\tilde{\tau}_{ij}$) and the filtered field $FF$ (the truth) are shown for the components of the symmetric stress tensor $\tilde{\tau_{ij}}$. 
        %The FF used for the input features to the models $M1$ and $M2$ are from the LES grid $36\times48\times36$ at $Re_\tau \sim 395$ listed in Tab. \ref{t2}. 
        }
        \label{fcorrtgt}
    \end{center}
\end{figure}

Perhaps, a better representation of the performance of the DL models is to show the correlation between the model predictions and the true SGS stresses from the $FF$. 
To compute the correlation, the following expression is used, 
\begin{equation}
\label{gama}
\gamma_{ij} = \frac{\langle(\tilde{\tau}^{FF}_{ij} - \langle \tilde{\tau}^{FF}_{ij} \rangle) (\tilde{\tau}^{m}_{ij} - \langle \tilde{\tau}^{m}_{ij} \rangle)\rangle} {\langle(\tilde{\tau}^{FF}_{ij} - \langle \tilde{\tau}^{FF}_{ij} \rangle)^2 \rangle^{\frac{1}{2}} \langle(\tilde{\tau}^{m}_{ij} - \langle \tilde{\tau}^{m}_{ij} \rangle)^2 \rangle^{\frac{1}{2}}}
\end{equation}
Here, $\tilde{\tau}^{FF}_{ij}$ is the true SGS stress for the component, and $\tilde{\tau}^{m}_{ij}$ is the SGS stress predicted by the models. 
A value of $\gamma = 1$ would indicate that the model predictions are same as the $FF$. 
On the other hand, a value of $\gamma=-1$ would indicate that the SGS stresses predicted by the model are anti-correlated to the true SGS stresses. 
Figure~\ref{fcorrtgt} presents $\gamma$ plotted as a function of $y+$ for all six components of $\boldsymbol{\tilde{\tau}}$ predicted by both models $M1$ and $M2$. 
Predictions made by $M2$ are clearly more correlated to the true SGS stresses close to the wall for all six components. 
$\gamma$ for $M1$ and $M2$ are similar away from the wall ($y+ > 100$), where the turbulence is more isotropic. 
$M2$ model yields a peak $\gamma \approx 0.5$ close to the wall for all six components. 
Only for the $\tilde{\tau}_{yz}$ component predictions by $M2$ are lower compared to $M1$ far away from the wall. 
Close to the wall, $M1$ provides negative values for $\gamma$ for all six components. 
This figure clearly demonstrates the superiority of the $M2$ model predictions, especially in predicting the anisotropic properties of the SGS stresses close to the wall.

\begin{figure}[!ht]
    \begin{center}
        \includegraphics[trim=0 0cm 0cm 0,clip,width=0.99\textwidth]{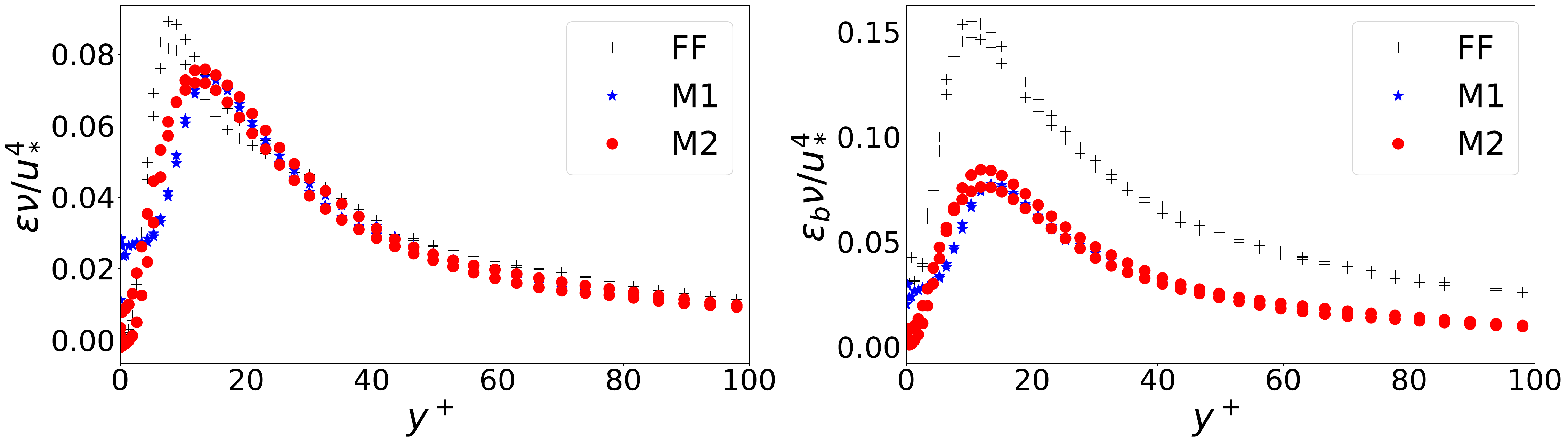}
        \caption{Testing performance of the $M1$ and $M2$ models compared with the filtered field (FF) on the training grid (dataset $D1$ in Tab. \ref{t2}): scaled mean SGS dissipation $\varepsilon$ (left), and scaled mean backscatter, $\varepsilon_b$ (right) due to the predicted SGS stresses. 
        %The FF used for the input features to the models $M1$ and $M2$ are from the LES grid $36\times48\times36$ at $Re_\tau \sim 395$ listed in Tab. \ref{t2}. 
        }
        \label{fdisstgt}
    \end{center}
\end{figure} 

The function of an SGS model is to provide a two-way interaction between the SGS flow and the resolved flow field. 
The first function is to dissipate energy from resolved flow field to the SGS flow. 
In the inertial range, dissipation is balanced by the production of energy, and hence, the dissipation provided to the grid-scale flow by an SGS model may be approximated by $\varepsilon \approx -\langle\tilde{\tau}_{ij}\tilde{S}_{ij}\rangle$. 
The other function is to model the forcing provided by the SGS flow to the resolved flow. 
This is called backscatter, which may be approximated as $\varepsilon_b \approx \frac{1}{2}(\varepsilon - \langle \varepsilon \rangle)$. 
$\varepsilon$ and $\varepsilon_b$ scaled by the viscous wall unit are plotted for the SGS stress predictions by the two models and compared with the true dissipation and backscatter obtained from the filtered fields in the left and right frames of Fig. \ref{fdisstgt}, respectively. 
Despite a significantly low model prediction variance, predicted $\varepsilon$ by both $M1$ and $M2$ are in good agreement with the true dissipation, marked as $FF$. 
The two models provide similar levels of dissipation. 
Especially, away from the wall, the SGS dissipation provided by both models is virtually the same as the true dissipation. 
On the other hand, the backscatter due to both models is low compared to the true backscatter denoted as $FF$. 
The location of the near-wall peak for both $\varepsilon$ and $\varepsilon_b$ is correctly captured by the two models. 
Although backscatter is necessary, it may also lead to numerical instability. 
Lower levels of backscatter predicted by these models indicate that both these models should be usable in a posteriori test, in an LES, implemented in a CFD program without numerical issues \cite{park2021toward}.

\subsection{Predictive performance on a different grid}

In this section, results from the $M1$ and $M2$ model predictions on the test dataset $D2$ listed in Tab. \ref{t2} are presented. 
The input features and outputs are generated with different filter sizes (smaller filter width) compared to the training data. 
However, the $Re$ is the same as that for the training flow.  

\begin{figure}[!ht]
    \begin{center}
        \includegraphics[trim=0 0cm 0cm 0,clip,width=0.99\textwidth]{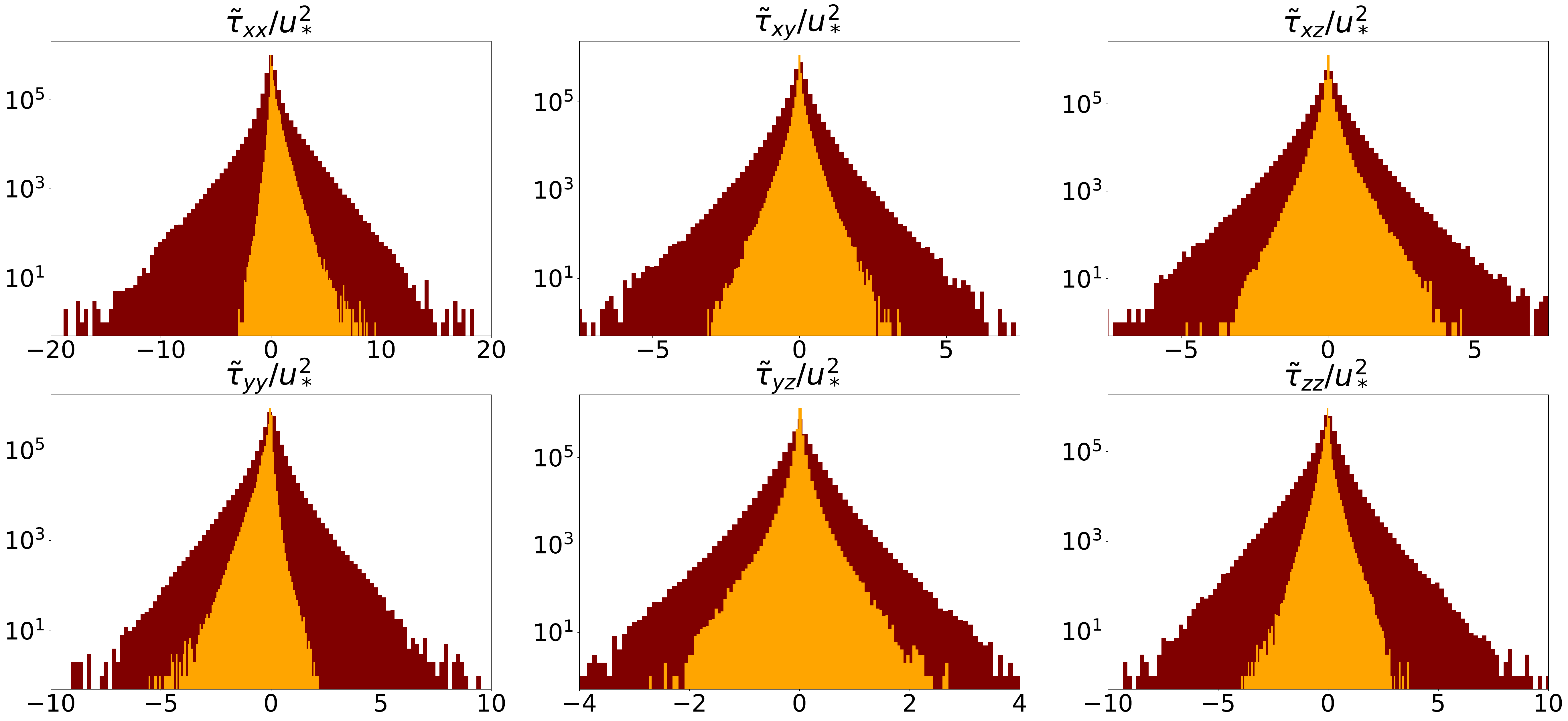}
        \caption{Testing performance of the $M1$ model on dataset $D2$ in Tab. \ref{t2}: frequency distribution of the predicted SGS stresses $\tilde{\tau}_{ij}$ (orange) and the filtered field $FF$ (maroon) are shown for the components of the symmetric stress tensor $\tilde{\tau_{ij}}$. 
        %The FF used for the input features to the model $M1$ are from the LES grid $48\times48\times48$ at $Re_\tau \sim 395$ listed in Tab. \ref{t2}. 
        }
        \label{fM1tg1}
    \end{center}
\end{figure}

\begin{figure}[!ht]
    \begin{center}
        \includegraphics[trim=0 0cm 0cm 0,clip,width=0.99\textwidth]{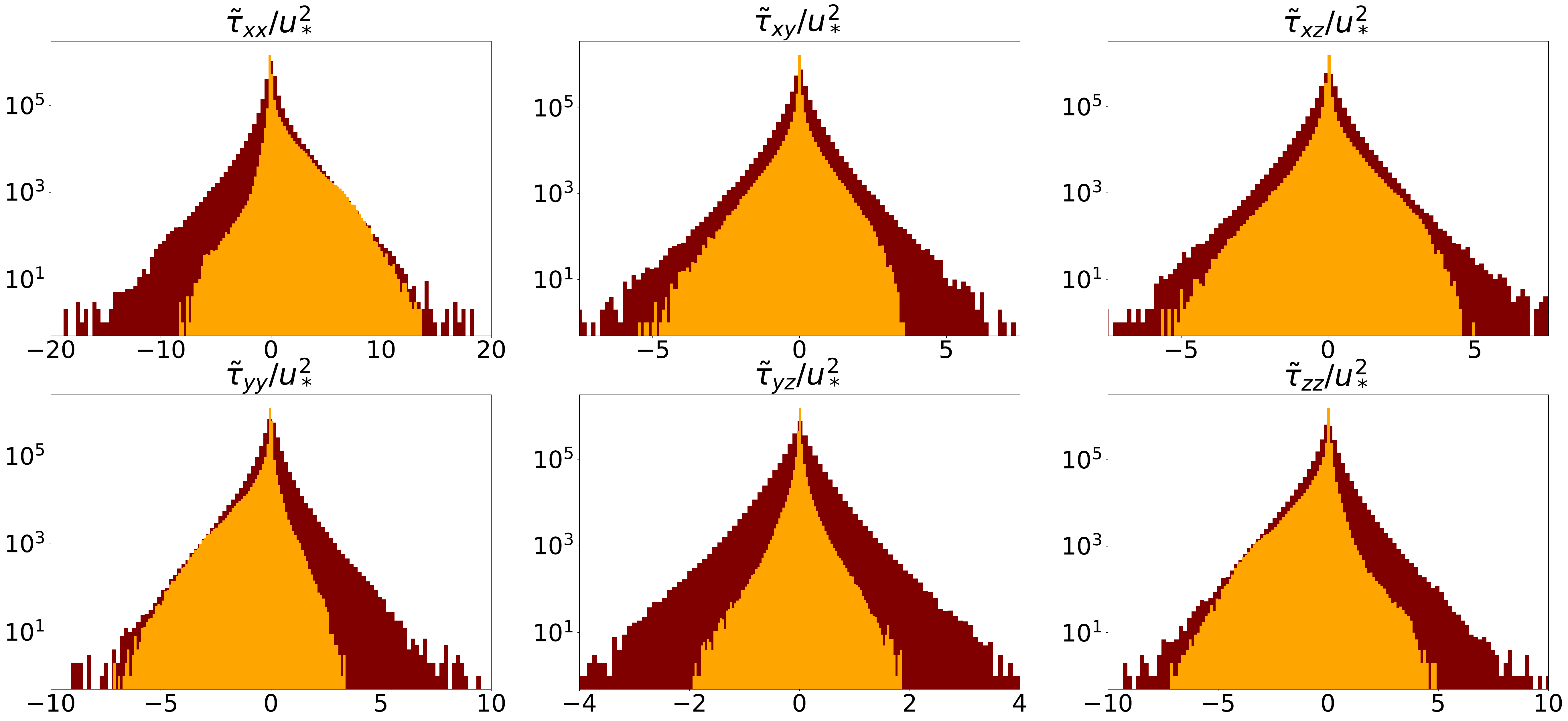}
        \caption{Testing performance of the $M2$ model on dataset $D2$ in Tab. \ref{t2}: frequency distribution of the predicted SGS stresses $\tilde{\tau}_{ij}$ (orange) and the filtered field $FF$ (maroon) are shown for the components of the symmetric stress tensor $\tilde{\tau_{ij}}$. 
        %The FF used for the input features to the model $M2$ are from the LES grid $48\times48\times48$ at $Re_\tau \sim 395$ listed in Tab. \ref{t2}. 
        }
        \label{fM2tg1}
    \end{center}
\end{figure}

Figures~\ref{fM1tg1} and \ref{fM2tg1} show the frequency distributions of the predictions provided by the $M1$ and $M2$ models for the $D2$ test dataset, respectively. 
The distributions for the model predictions are colored orange and the distributions for the true SGS stresses are colored maroon. 
From the comparison it becomes clear that the stresses predicted by $M1$ have lower variance compared to the $M2$ model predicted stresses. 
These plots also show that the true values for the $\tilde{\tau}_{xx}$ component are larger than those of the other five components. 
Possibly because of this property, values of $\tilde{\tau}_{xx}$ predicted by $M1$ seem to have notably lower variance than the other five components. 
In comparison, model $M2$ is able to predict the tail ends of the distributions for all the six components better than $M1$. 
Predictions made by $M2$ are in good agreement with the true SGS stresses for this dataset as well. 
Overall, even on a significantly different grid than the training grid, the performance of model $M2$ is satisfactory. 
However, there is notable asymmetry for the model predicted distributions compared to the obviously symmetric distributions observed for the true SGS stresses.

\begin{figure}[!ht]
    \begin{center}
        \includegraphics[trim=0 0cm 0cm 0,clip,width=0.99\textwidth]{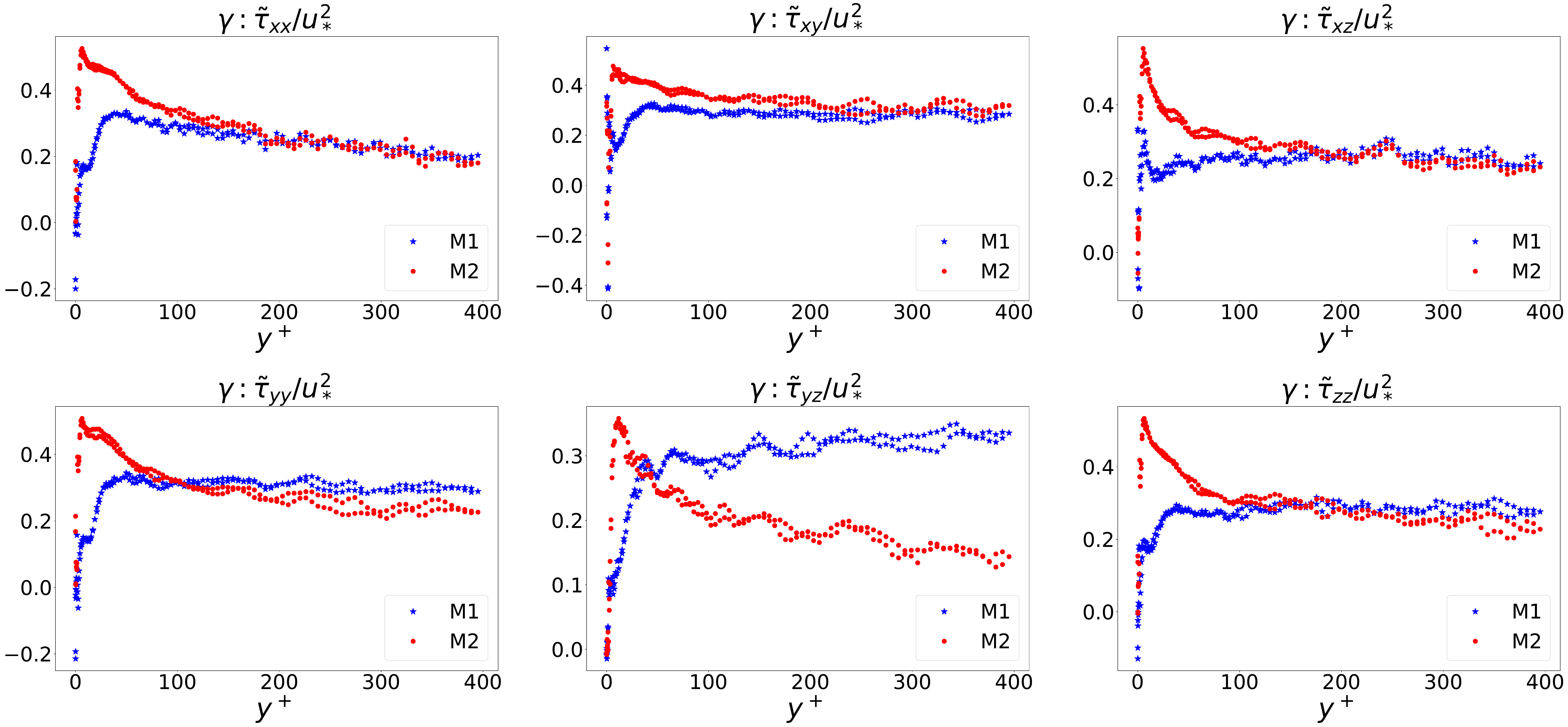}
        \caption{Comparison of the testing performance of the $M1$ and $M2$ models on dataset $D2$ in Tab. \ref{t2}: correlation coefficient ($\gamma$) computed between the predicted SGS stresses ($\tilde{\tau}_{ij}$) and the filtered field $FF$ (the truth) are shown for the components of the symmetric stress tensor $\tilde{\tau_{ij}}$. 
        %The FF used for the input features to the models $M1$ and $M2$ are from the LES grid $48\times48\times48$ at $Re_\tau \sim 395$ listed in Tab. \ref{t2}. 
        }
        \label{fcorrtg1}
    \end{center}
\end{figure}

The correlation coefficients ($\gamma$) between the model predicted and the true SGS stresses are shown as a function of the wall-normal coordinate in Fig. \ref{fcorrtg1}. 
%As was the case for the dataset $D1$ in Fig. \ref{fcorrtgt}, $M2$ predicted SGS turbulent stresses for dataset $D2$ are remarkably correlated to the true SGS stresses close to the wall. 
As was the case for dataset $D1$ in Fig. \ref{fcorrtgt}, $M2$ predicted SGS stresses are most accurate in the near-wall layer with a peak correlation value of $\gamma \sim 0.5$ for all the components. 
On the other hand, $M1$ predicted stresses are relatively more correlated to the true stresses away from the wall. 
Close to the wall, $M1$ predicted stresses are negatively correlated to the true stresses for all six components. 
Except for the $\tilde{\tau}_{xy}$ component away from the wall, $M2$ predicted stresses are in better overall agreement with the true SGS stresses for all the components of $\boldsymbol{\tilde{\tau}}$. 
Overall, the predictions by both models on datasets $D1$ and $D2$ are consistent. 
The superior performance of the $M2$ model is retained even when the filter width for the test dataset is significantly reduced compared to the training grid.

\begin{figure}[!ht]
    \begin{center}
        \includegraphics[trim=0 0cm 0cm 0,clip,width=0.99\textwidth]{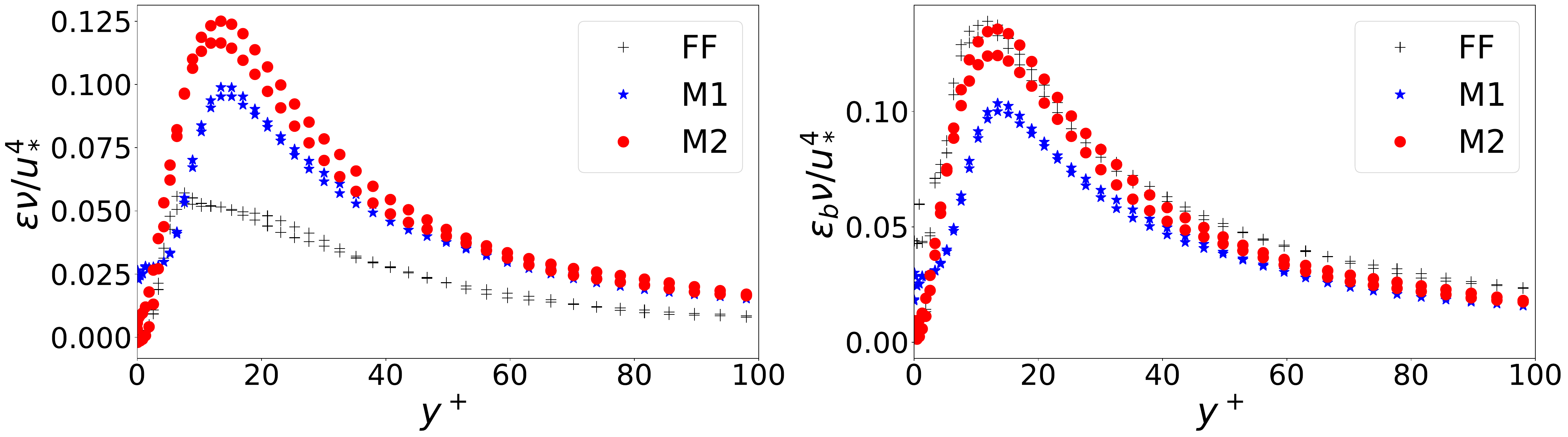}
        \caption{Testing performance of the $M1$ and $M2$ models compared with the filtered field (FF) on the test grid (dataset $D2$ in Tab. \ref{t2}): scaled mean SGS dissipation $\varepsilon$ (left), and scaled mean backscatter, $\varepsilon_b$ (right) due to the predicted SGS stresses.  
        %The FF used for the input features to the models $M1$ and $M2$ are from the LES grid $48\times48\times48$ at $Re_\tau \sim 395$ listed in Tab. \ref{t2}.  
        }
        \label{fdisstg1}
    \end{center}
\end{figure}

Wall-normal profiles of dissipation, $\varepsilon$, and backscatter provided by models $M1$ and $M2$ for the dataset $D2$ are shown in the left and right frames of Fig. \ref{fdisstg1}, and compared with the profiles of true dissipation and backscatter obtained from $FF$. 
Compared to the the truth, $\varepsilon$ is overpredicted by both models; dissipation provided by $M2$ is higher than $M1$ close to the wall. 
However, the characteristics of the profiles are similar to those of the profiles of true dissipation close to the wall. 
Both models provide similar levels of dissipation away from the wall, for $y+ \ge 50$. 
On the other hand, $\varepsilon_b$ profile for the $M2$ model predictions are in very good agreement with the truth, especially close to the wall. 
Away from the wall, backscatter is only slightly underpredicted by $M2$. 
On the other hand, $\varepsilon_b$ predicted by $M1$ is lower than the true $\varepsilon_b$ for the $FF$, especially close to the wall, although the location of the near-wall peak is in agreement. 
The deviation from the true backscatter reduces away from the wall for this model.

\subsection{Predictive performance at a different Reynolds number}

In this section, we present results for the model performance on dataset $D3$ listed in Tab. \ref{t2}. 
For this dataset, the Reynolds number is $Re_\tau\approx 590$ compared to the Reynolds number for the training flow, $Re_\tau \approx 395$, although the filter widths in all three spatial directions for this test dataset and training data are similar.

\begin{figure}[!ht]
    \begin{center}
        \includegraphics[trim=0 0cm 0cm 0,clip,width=0.99\textwidth]{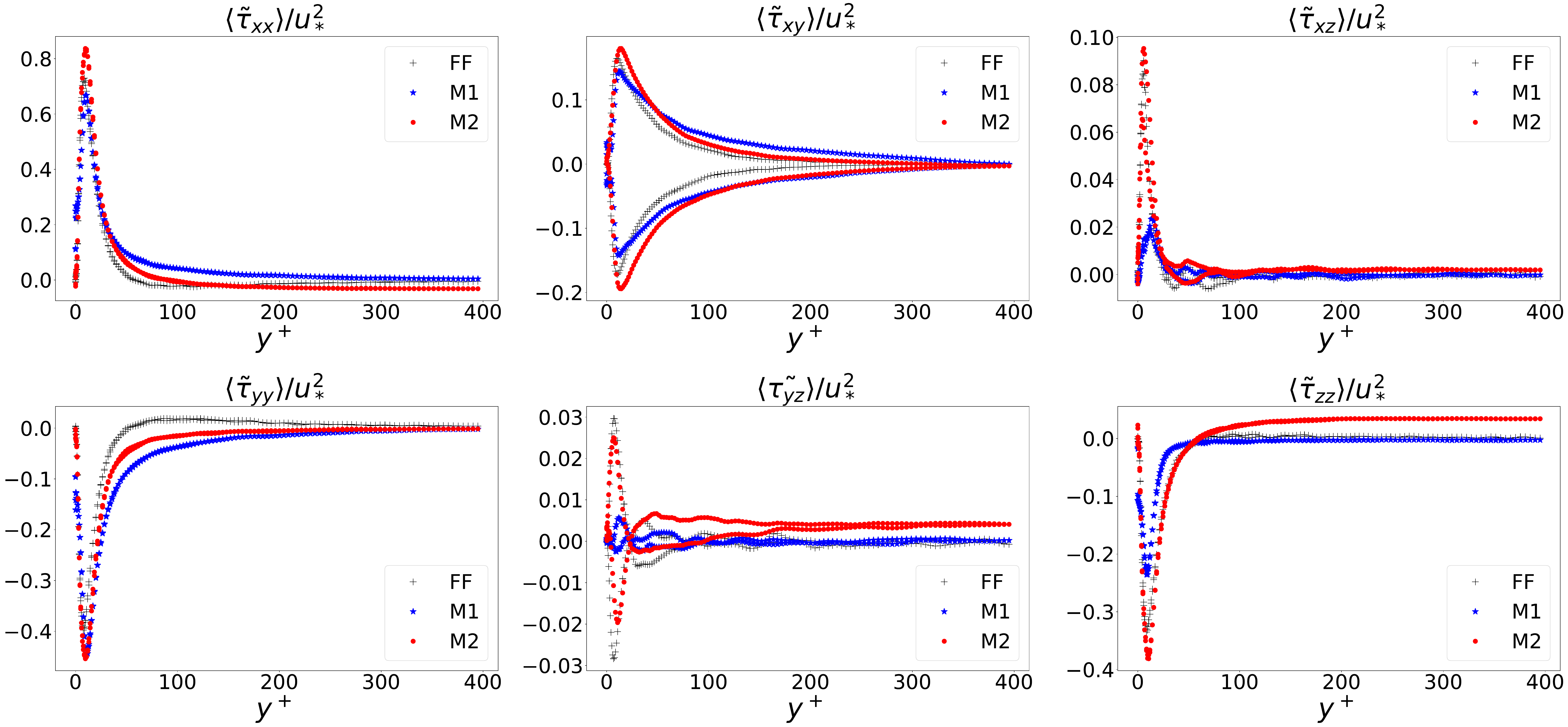}
        \caption{Testing performance of the $M1$ and $M2$ models compared with the filtered field (FF) on a testing grid at a different $Re$ (dataset $D3$ in Tab. \ref{t2}): scaled mean SGS stresses ($\langle \tilde{\tau_{ij}} \rangle$) are shown for the components of the symmetric stress tensor $\tilde{\tau_{ij}}$. 
        %The FF used for the input features to the models $M1$ and $M2$ are from the LES grid $54\times72\times54$ at $Re_\tau \sim 590$ listed in Tab. \ref{t2}. 
        }
        \label{fmtg2}
    \end{center}
\end{figure} 

Figure~\ref{fmtg2} presents the wall-normal profiles of the horizontally and time averaged SGS stresses. 
The predictions by the models $M1$ and $M2$ are compared with the true stresses obtained from $FF$. 
As was the case for dataset $D1$ in Fig. \ref{fmtgt}, mean SGS stresses predicted by model $M2$ are in excellent agreement with the truth. 
In particular, the near-wall peaks in the profiles for all six components are accurately captured by $M2$. 
On the other hand, model $M1$ slightly underpredicts the the near-wall peak. 
Except for the $\tilde{\tau}_{yz}$ and $\tilde{\tau}_{zz}$ components, away from the wall, predictions by both models are in agreement with the true stresses. 
For these two components, $M2$ slightly overpredicts the stresses in the free stream. 
Despite the dataset being for a significantly higher $Re$, both models were able to extrapolate successfully. 
Overall, the near-wall anisotropy is better captured by $M2$. 

\begin{figure}[!ht]
    \begin{center}
        \includegraphics[trim=0 0cm 0cm 0,clip,width=0.99\textwidth]{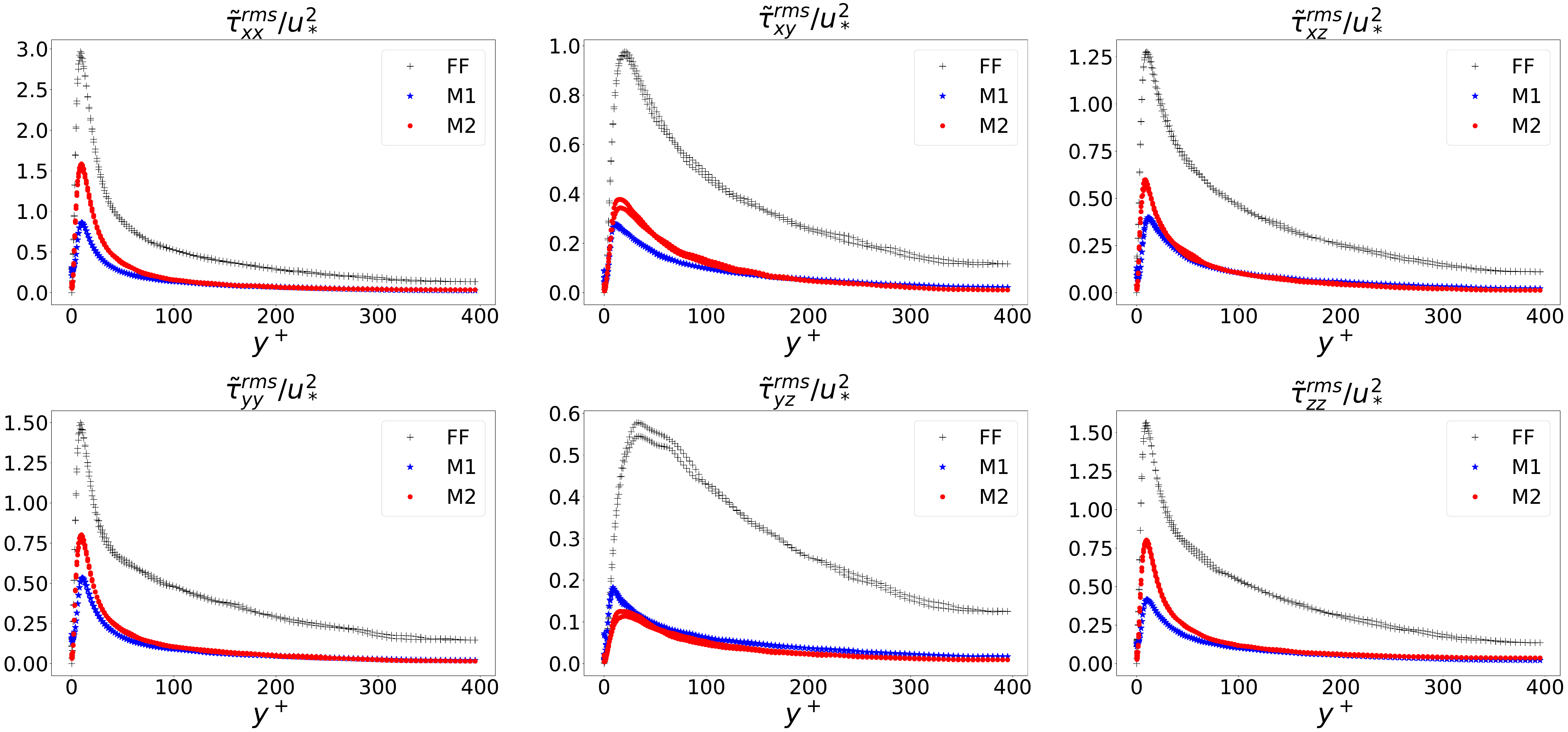}
        \caption{Testing performance of the $M1$ and $M2$ models compared with the filtered field (FF) on a testing grid at a different $Re$ (dataset $D3$ in Tab. \ref{t2}): scaled $r.m.s.$ SGS stresses ($\tilde{\tau}^{rms}_{ij}$) are shown for the components of the symmetric stress tensor $\tilde{\tau_{ij}}$. 
        %The FF used for the input features to the models $M1$ and $M2$ are from the LES grid $54\times72\times54$ at $Re_\tau \sim 590$ listed in Tab. \ref{t2}. 
        }
        \label{frmstg2}
    \end{center}
\end{figure} 

The profiles of the $r.m.s.$ values of the model predicted stresses averaged in the statistically homogeneous directions and time snapshots are compared with the profiles of the $r.m.s.$ true SGS stresses in Fig. \ref{frmstg2}. 
The variance of the model predictions for all components for both models are significantly lower than that for the true stresses denoted as $FF$. 
Variance for the $M2$ model predictions are larger than those for the $M1$ model predictions. 
The location of the near-wall peak in variance for the different components is correctly captured by both models. 
This figure again demonstrates the superiority of approach $A2$ compared to $A1$ towards obtaining a more accurate model for predicting the SGS stresses. 
As previously noted, the spatial gradients of the predicted SGS stresses are more important than their absolute values. 
This figure is presented herein from a typical DL model testing context. 

\begin{figure}[!ht]
    \begin{center}
        \includegraphics[trim=0 0cm 0cm 0,clip,width=0.99\textwidth]{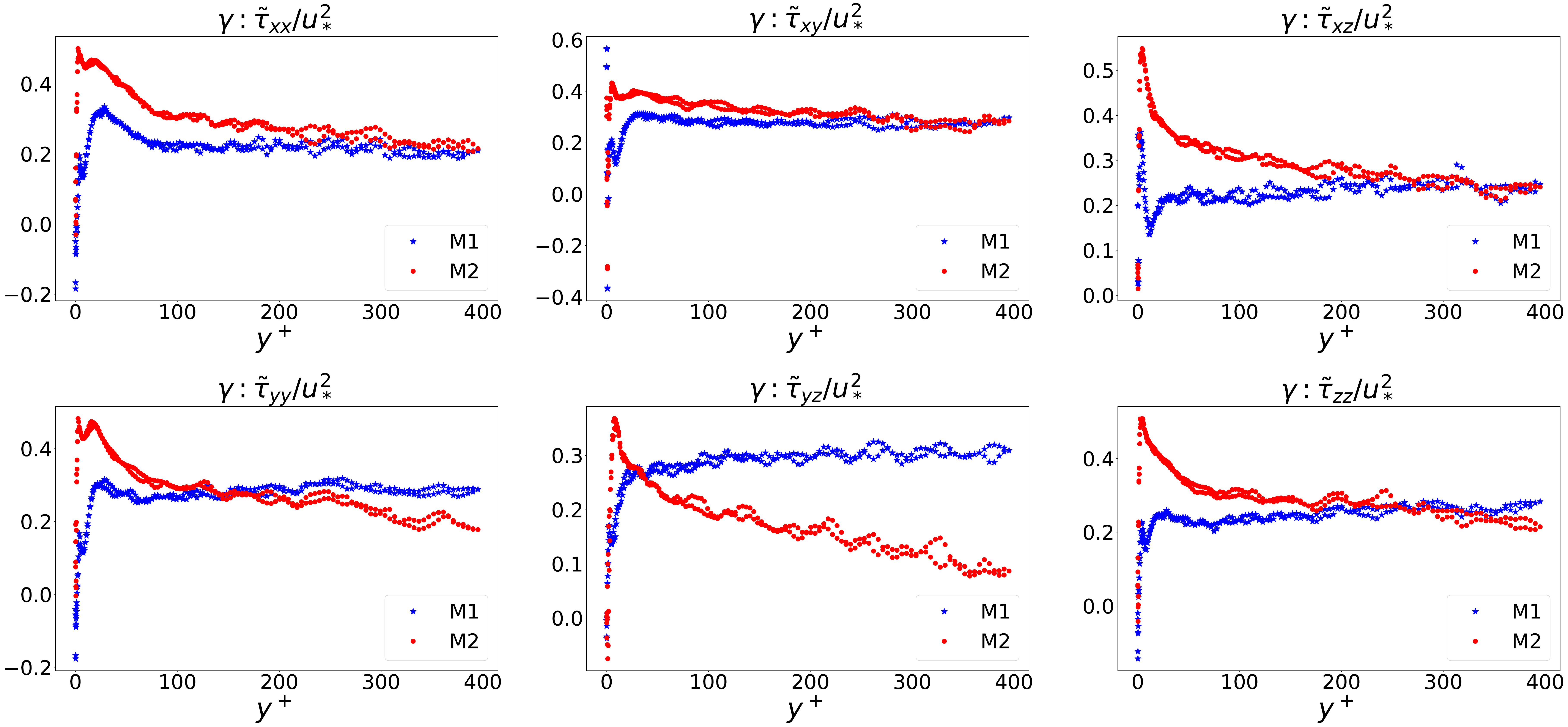}
        \caption{Comparison of the testing performance of the $M1$ and $M2$ models at a different $Re$ (dataset $D3$ in Tab. \ref{t2}): correlation coefficient ($\gamma$) computed between the predicted SGS stresses ($\tilde{\tau}_{ij}$) and the filtered field $FF$ (the truth) are shown for the components of the symmetric stress tensor $\tilde{\tau_{ij}}$. 
        %The FF used for the input features to the models $M1$ and $M2$ are from the LES grid $54\times72\times54$ at $Re_\tau \sim 590$ listed in Tab. \ref{t2}. 
        }
        \label{fcorrtg2}
    \end{center}
\end{figure}

Correlation coefficients ($\gamma$) are shown for the six components of the SGS stresses for the $M1$ and $M2$ model predictions in Fig. \ref{fcorrtg2}. 
Again, except for the component $\tilde{\tau}_{yz}$, the predictions made by model $M2$ are more accurate, both close to the wall, and away from it, compared to model $M1$ even for this higher $Re$ dataset. 
Similar to its performance on datasets $D1$ and $D2$, for dataset $D3$, model $M2$ model is able to learn and predict the near-wall anisotropy more accurately. 
Also similar to the previously shown results for $\gamma$, accuracy of $M2$ reduces away from the wall. 
Similar prediction accuracy of the $M1$ model on all three datasets, especially away from the wall, indicate that the TBNN architecture invented to emulate the effective viscosity hypothesis model form is more suitable for modeling  isotropic turbulence. 
On the other hand, despite only embedding the Galilean invariance property, model $M2$ is more accurate at predicting the SGS stresses in the anisotropic near-wall regions. 
It also shows similar accuracy when predicting stresses away from the wall where turbulence is more isotropic in nature. 

\begin{figure}[!ht]
    \begin{center}
        \includegraphics[trim=0 0cm 0cm 0,clip,width=0.99\textwidth]{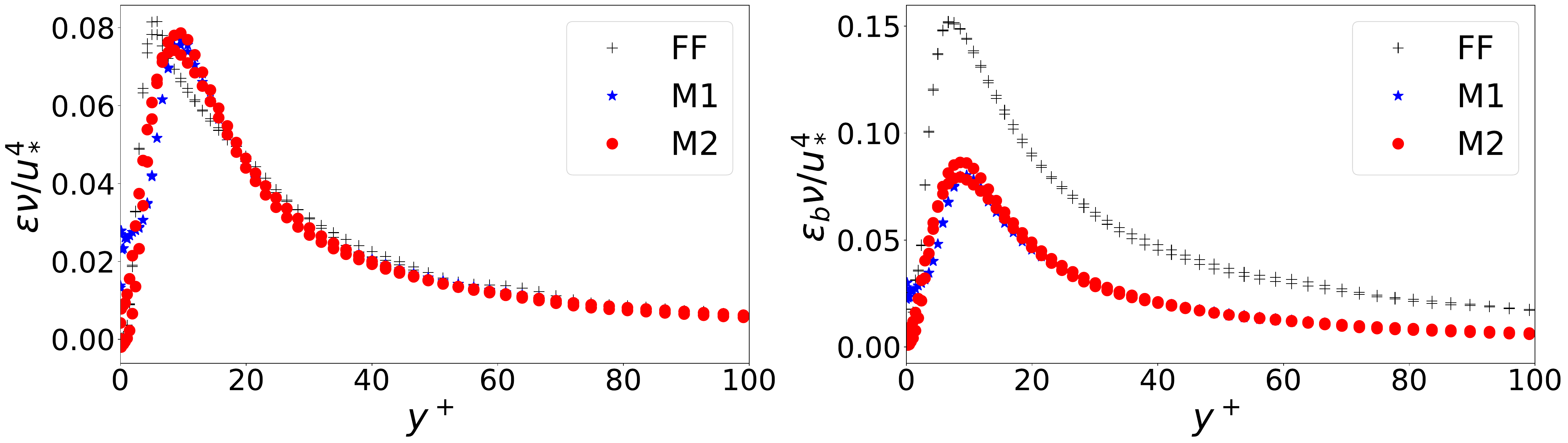}
        \caption{Testing performance of the $M1$ and $M2$ models compared with the filtered field (FF) on a testing grid at a different $Re$ (dataset $D3$ in Tab. \ref{t2}): scaled mean SGS dissipation $\varepsilon$ (left), and scaled mean backscatter, $\varepsilon_b$ (right) due to the predicted SGS stresses. 
        %The FF used for the input features to the models $M1$ and $M2$ are from the LES grid $54\times72\times54$ at $Re_\tau \sim 590$ listed in Tab. \ref{t2}.  
        }
        \label{fdisstg2}
    \end{center}
\end{figure}

The left and right frames of Fig. \ref{fdisstg2} show the comparisons of the wall-normal profiles of true $\varepsilon$ and $\varepsilon_b$, respectively, obtained from the $FF$, and those obtained from the predictions made by models $M1$ and $M2$. 
As was the case for dataset $D1$ at a lower $Re$, the profiles of $\varepsilon$ for both models are in excellent agreement with the true $\varepsilon$ from the $FF$. 
Model $M2$ is more accurate at predicting the profile of $\varepsilon$. 
The near-wall peak for dissipation is predicted at a slightly higher wall-normal location by both models. 
The predictions by the models away from the wall are also same. 
However, both models underpredict the backscatter, $\varepsilon_b$ in the right frame of Fig. \ref{fdisstg2}, especially close to the wall. 
The deviation from the truth reduces away from the wall. 
The backscatter estimates provided by the models are very similar for this higher $Re$ case.

\section{Discussion \& Conclusions}\label{sec5}

In this paper, two families of SGS turbulence models for LES are developed based on the powerful DL framework. 
These two families of NNs are distinct in their architecture. 
Although the input features for these NN models are based on the same set of variables from the resolved flow field in an LES, because of the difference in their architecture, they have different feature extraction capabilities. 
The NN architecture of the first family of models ($A1$) embeds the analytical form of the generalized effective-viscosity hypothesis \citep{pope1975more, lund1993parameterization}, and is therefore, able to embed Galilean, rotational and reflectional invariances. 
This innovative NN architecture has separate scalar and tensor input layers \citep{ling2016reynolds}. 
The inputs to the invariance input layer to these family of models are the six scalar invariants based on the strain rate ($\boldsymbol{\tilde{S}}$) and rotation rate ($\boldsymbol{\tilde{R}}$) tensors, and their higher powers. 
The tensor input layer inputs the integrity basis tensors, and feature extraction from these tensors are implicit in nature; only component-wise feature extraction is possible by optimization of the trainable parameters via the minimization of the loss function; the cross-component feature extraction is inhibited in this NN architecture (see Fig. \ref{fa1}). 
The models developed using the second approach ($A2$) on the other hand are only invariant to the Galilean transformation, and only usable in wall-bounded turbulence, where the wall provides an appropriate reference frame. 
The architecture for this set of models is simpler. 
Apart from the six invariant inputs, the inputs to the models developed in this approach include all six independent components of each of the five integrity basis tensors; despite its simpler architecture, it allows for feature extraction from all the cross-component terms of the basis tensors (see Fig. \ref{fa2}).

The NN-SGS models are trained and tested on data generated by explicitly filtering flow fields generated in DNSs of channel flow at different Reynolds numbers (Tab. \ref{t1}). 
The sharp spectral cutoff filter is used for explicit filtering to ensure that the Galilean invariance is automatically satisfied for the filtered SGS stresses \citep{speziale1985galilean} (see Fig. \ref{fFilt2}). 
Filtering is performed in all three spatial directions. 
Three datasets are generated; one of these is used for both training and testing, while the other two are only used for testing the trained models' performance on features generated with (i) a different filter width compared to the training data, and (ii) with similar filter widths, but at a different $Re$ than the training dataset (Tab. \ref{t2}). 
For data scaling, fluid dynamic variables were found to augment the models' ability to fit the training data better than the standard scalings, such as, normalization and standardization, typically used in the field of DL. 
The hyper-parameter space related to the NN architecture was extensively searched to find the optimal models (see Tabs. \ref{tA1} and \ref{tA2}). 
For fairness in comparison, the hyper-parameters were chosen, so that two models from either family with the same number of hidden layers would also have very similar number of trainable parameters. 
However, the dependence of the model performance is relatively insensitive to the hyper-parameters once a level of complexity is incorporated in the NN architecture for both sets of models. 
In our experience of training the NN models, the Leaky $ReLU$ activation is inappropriate for the DL-SGS models considered herein, as these may lead to sudden discrete jumps in the values of the model parameters (see Figs. \ref{feA1} and \ref{feA2}). 
The sigmoid activation was found to provide consistent performance over training, validation and test datasets for both sets of models.

Following a systematic procedure, only one model was selected out of the 36 models trained in each approach. 
These two models, $M1$ and $M2$, representatives of the two approaches considered in this study, were extensively tested on the three test datasets to explore the viability of these two approaches in a posteriori testing. 
Both statistical and fluid dynamic parameters were considered to gauge the performance of the models. 
Within the statistical parameters, mean, variance and correlation coefficients between the prediction and the truth were examined. 
As the DL-SGS models are supposed to provide two-way coupling between the resolved and the SGS fields, dissipation and backscatter provided by these models were also tested. 
Overall, for all three datasets, the $M2$ model was found to provide notably superior performance. 
Except for one of the six predicted components (this component generally has the smallest magnitude out of the six predicted components), model $M2$ provided better mean and variance profiles (Figs. \ref{fmtgt}, \ref{frmstgt}, \ref{fM2tg1}, \ref{fmtg2}, and \ref{frmstg2}), and more accurate predictions quantified by the correlation coefficient between the prediction and the truth (Figs. \ref{fcorrtgt}, \ref{fcorrtg1}, \ref{fcorrtg2}). 
The predictions by model $M1$ are more accurate away from the wall where the turbulence is more isotropic in nature. 
Compared to $M1$, model $M2$ was able to provide significantly improved predictions close to the wall, implying, this model is able to learn the near-wall anisotropic nature of turbulence. 
Additionally, $M2$ also performed slightlu better in predicting the fluid dynamic parameters, such as, dissipation and backscatter. 
Although, the wall-normal profiles of mean dissipation were satisfactory for both models (Figs. \ref{fdisstgt}, \ref{fdisstg1}, \ref{fdisstg2}), the backscatter was underpredicted by both.

In spite of using a simpler network architecture, model $M2$ was found to provide significantly more accurate predictions. 
Hence, it may be concluded that, in general, approach $A2$ would yield more accurate models. 
It must be noted however, that the input features for the NN-family $A2$ are also motivated by the same physical considerations as the NN-family $A1$; the only difference being the embedding of the analytical model form in Eq. \ref{Pope} in the $A1$ model architecture, which in turn incorporates several invariance properties, the Galilean, rotational, and reflectional invariances. 
Perhaps as expected, utilization of the information available in the cross-components of the basis tensors results in better feature extraction from the underlying data by the $A2$ NN models. 
The performance tests presented in this work are only a priori in nature. 
The present study motivates the use of these models in actual LESs. 
The current focus of research in this direction is in a posteriori testing of these models in actual LESs, which will be the subject of a future paper. 
Furthermore, it could be possible to use the principal invariants of the integrity basis tensors as inputs to a NN that could result in another approach reinstating the invariance properties not incorporated in the $A2$ family of models. 
Whether this alternative approach could yield predictions at least as accurate as the $A2$ family of models is an open question.

\iffalse
\begin{acknowledgments}
Helpful comments by Dr. Emil Simiu of the National Institute of Standards and Technology are acknowledged with thanks. 
\end{acknowledgments}
\fi

%The data that support the findings of this study are available from the corresponding author upon reasonable request. 

\pagebreak
% Create the reference section using BibTeX:
\bibliography{reference}
\end{document}